\documentclass[manuscript]{aastex}
\usepackage[varg]{txfonts}
\usepackage{graphicx,subfigure}
\usepackage[]{natbib}
\shorttitle{Lags in GX 5-1}
\shortauthors{Sriram et al.}

\begin{document}
\title{Anti-correlated time lags in the Z source GX 5-1: Possible evidence for a truncated accretion disk}
%\subtitle{I. Proof for truncated disk}

\author{K. Sriram, C. S. Choi}
\affil{Korea Astronomy and Space Science Institute, Daejeon 305-348, Republic of Korea}

\email{astrosriram@yahoo.co.in}
\author{A. R. Rao}
          
\affil{Tata Institute of Fundamental Research, Mumbai 400005, India}

\begin{abstract}
We investigate the nature of the inner accretion disk in the neutron star source GX 5-1
by making a detailed study of time lags between X-rays of different energies.
Using the cross-correlation analysis,
we found anti-correlated hard and soft time lags of the order of a few tens to a few
hundred seconds and the corresponding intensity states 
were mostly the horizontal branch (HB) and upper normal branch (NB). 
The model independent and dependent spectral analysis showed 
that during these time lags the structure of accretion disk significantly varied. 
Both eastern and western approaches were used to unfold the X-ray continuum 
and systematic changes were observed in soft and hard spectral components. 
These changes along with a systematic shift in the frequency of quasi-periodic oscillations (QPOs)
made it substantially evident that the geometry of the accretion disk is truncated. 
Simultaneous energy spectral and power density spectral study shows that the 
production of the horizontal branch oscillations (HBOs) are closely related to the 
Comptonizing region rather than the disk component in the accretion disk. 
We found that as the HBO frequency decreases from the hard apex to upper HB, the disk temperature 
increases along with an increase in the coronal temperature which is in sharp contrast with the changes 
found in black hole binaries where the decrease in QPO frequency is accompanied by a decrease in the disk 
temperature and a simultaneous increase in the coronal temperature. We discuss the
results in the context of re-condensation of coronal material in the inner region of the disk. 
\end{abstract}
    

\section{Introduction}

Neutron star low mass X-ray binaries (LMXBs) can be divided into two broad classes viz. Z-type and atoll-type,
based on their color-color diagram (CCD) or hardness-intensity diagram (HID), where the Z-type sources
have luminosities close to the Eddington limit (van der Klis 2006).
It has been suggested that the specific positions on the CCD or HID tracks indicate a different physical configuration
of an accretion disk or radiation process (Hasinger \& van der Klis 1989; Hasinger et al. 1990). 
The Z-type sources traverse a `Z'-shaped path on HID which consists of three main branches viz. horizontal branch (HB), 
normal branch (NB) and flaring branch (FB), where NB connects to the HB and NB. 

The Z-type sources can be
subdivided into Sco X-1-like (e.g., GX 17+2 and GX 349+2) and Cyg X-2-like (e.g., GX 5-1 and GX 340+0)
(Kuulkers et al. 1994). The Sco X-1-like sources exhibit a vertical HB whereas the latter sources exhibit
a horizontal HB (Kuulkers et al. 1997). It was suggested that the difference is due either to the magnetic field 
strength of the neutron star (Psaltis et al. 1995) or to the inclination angle to the line of sight
(Hasinger et al. 1989; Hasinger et al. 1990; Kuulkers et al. 1994).
However, recent studies on the source XTE J1701-462 ruled out both the possibilities (Homan et al. 2010)
because this source showed both Cyg X-2- and Sco X-1-like properties (Homan et al. 2007). 
The physical parameter which causes the motion along the Z track on HID was considered to be the mass accretion rate, 
where the rate increases from the HB to NB and peaks at the FB.
It was also proposed that the accretion rate increases
from the FB to HB, which is exactly the opposite case (Jackson et al. 2009). 
A detailed study on the source XTE J1701-462 suggested that an instability may cause the Z track rather than the
accretion rate because in the study the accretion rate was nearly constant along the tracks
(Lin et al. 2009; Homan et al. 2010). 

There has been no consensus on the origin of soft X-rays for Z-type sources. The neutron star surface
or the Keplerian component of the disk could be the possible site.
The hard component in the X-ray spectrum is considered to originate from the Comptonization of low energy
photons in a hot corona, whose location and structure are not yet properly understood. There are three  
approaches to model the X-ray spectra of Z-type sources, western, eastern and hybrid model (Lin et al. 2009).
The western model (White et al. 1986) consists of a high energy (or power-law) component which is radiating by 
the Comptonization from the inner region of the disk and a soft/black body component radiating from the
surface or close to the surface of neutron star (NS).
In the eastern model (Mitsuda et al. 1984), the soft component is assumed to originate from the accretion
disk which radiates as a multi-temperature black body and the hard component arises from the Compton cloud/corona
assumed to be located at the inner region of the disk or close to the NS surface (Di Salvo et al. 2000,
2001; Agrawal \& Misra 2009). Generally the shape of Compton cloud/corona is considered to be quasi-spherical
(Zdziarski et al. 1996; Done et al. 2007) but the studies based on dipping LMXBs show that corona is more
likely to be an extended body over the disk (Church \& Balucinska-Church 2001; Jackson et al. 2009).
In black hole X-ray binaries (BHXBs), the unfolded spectrum to some extent resembles the eastern
model (i.e. soft component from the disk and hard component from the Compton cloud or at the base of jet)
and a geometry of truncated accretion disk is often implemented to explain the hard spectral state in BHXBs 
(see Done et al. 2007 for a review).
Both western and eastern models, however, are generally found to be successful in explaining
the spectra of neutron star LMXBs (Hasinger et al. 1990) but for some sources hybrid model is preferred 
over the former models (Lin et al. 2007; Ng et al. 2010).
Generally during an outburst in BHXBs, the spectrum is dominated by the hard component (along with a strong
radio emission) and disappears as the disk moves toward the last stable orbit (Remillard \& McClintock 2006;
Done et al. 2007). Similar phenomenon is often observed in Z-type sources where the hard X-ray tail and radio
emission are observed at HB and fade out when the source is at FB (Di Salvo et al. 2001; Migliari et al. 2007;
Lin et al. 2009).  

The signature of the Comptonization process can be obtained from cross-spectral studies. 
Both hard and soft lags of the order of a few micro/milli seconds exist, which are considered to be the possible
time taken by the soft photons to reprocess as hard photons in the hot corona (van der Klis et al. 1987). Nobili et al. 
(2000) suggested an inhomogeneous Compton cloud model to explain the soft lags. 
Such kind of lags were seen in GX 5-1 and the detection of lags were
related to the specific position on the Z-track (Vaughan et al. 1999; Qu et al. 2004). 
Anti-correlated hard or soft lags greater than a few tens of seconds,
however, pertains to the pivoting of the spectrum, where anti-correlation means that an intensity variation
is opposite between soft and hard X-rays and hard/soft lag means that hard/soft X-ray photons are relatively
delayed in their arrivals at the observer.
On the basis of a simple Comptonization model, Zdziarski et al. (2002) found that the spectral pivoting 
(and hence the anti-correlation) in the BHXB Cyg X-1 can be explained by assuming
a variable seed photon flux for the Comptonization process.
Any delays in such pivoting process will shed light on the physical processes responsible for the
generation of the seed photons and/or the re-distribution of geometry of the corona.

Anti-correlated X-ray lags of a few hundred seconds were often seen in some BHXBs (Choudhury et al. 2005;
Sriram et al. 2007; Sriram et al. 2009; Sriram et al. 2010) and based on the observed spectral pivoting the
lags were explained in the context of truncated accretion disk scenario. 
The simulation carried out by Sriram et al. (2010) rules out the spurious nature of the observed anti-correlation. 
So far such anti-correlated hard and soft lags (a few 10-100 s) were detected only for
one neutron star source, Cyg X-2, and most of the corresponding observations were located at the HB and upper
NB on the Z-track (Lei et al. 2008). Detection of such lags in neutron star LMXBs is important because it indirectly helps to know the radiative
and geometrical structure of the accretion disk (which is poorly known in the case of NS LMXBs).
The aforementioned arguments are the motivation of our study and we carry out a systematic timing analysis, 
cross-correlation analysis and the study of power density spectrum (PDS), along with the spectral analysis
for the source GX 5-1.              

GX 5-1 is the second brightest persistent LMXB located at the Sagittarius region (Bradt et al. 1968; Fisher et al.
1968) and it is classified as a Z-type source (Hasinger \& van der Klis 1989). The horizontal branch oscillations
(HBOs) with frequencies between 13-50 Hz were discovered by van der Klis et al. (1985) and the normal branch
oscillations (NBOs) with frequencies $\sim$6 Hz were found by Lewin et al. (1992) in GX 5-1. 
Recently, Sriram et al. (2011b) found that coupled HBO and NBO variations in the source can be interpreted as
the physically modified inner disk front.
The kHz quasi-periodic oscillations (QPOs) with frequencies ranging from 200-800 Hz have been detected at the HB
and upper part of NB (Wijnands et al. 1998), in which the QPO peaks are not constantly separated (Jonker et al. 2002).
However, no QPOs have been detected at the FB.  
GX 5-1 shows an inherent hard X-ray tail (power-law index of $\Gamma\sim$1.8) and a Gaussian-like residuals feature 
which is found at $\sim$10 keV over the eastern and western models (Asai et al. 1994).
The existence of jets in GX 5-1 was inferred from the detection of radio emission 
(Penninx et al. 1989; Tan et al. 1992; Fender \& Hendry 2000).  
Emission in infrared wavelength was also detected (Jonker et al. 2000). 

\section{Data Reduction and Analysis}

We analyzed the public archival data obtained with the Proportional Counter Array (PCA) on board the
{\it Rossi}-XTE satellite. The PCA consists of five identical proportional counter units which typically cover
the energy range of 2 -- 60 keV (Jahoda et al. 2006). Among the available 183 data sets for GX 5-1,
we selected 74 data sets that have exposures greater than 2000 s.
For spectral analysis, we extracted energy spectra from PCU2 which is the best calibrated among
the PCUs and added 0.5\% systematic errors to account for the calibration uncertainties.
HEASOFT v6.8 software was used to reduce the raw data of GX 5-1 and unfolded spectra were obtained
using the spectral models in XSPEC v12.5.0 (Arnaud 1996).

To obtain the HID, we defined the hardness ratio as the ratio of background subtracted source counts in
8.7 -- 19.7 keV to those of 6.2 -- 8.7 keV and calculated the intensity in a bin size of 128 s (count rate)
for the energy range 2.0 -- 19.7 keV (Qu et al. 2004).
The channels were carefully selected based on the different RXTE gain epochs and the gain-change effect was corrected.
For the cross-correlation analysis, the background subtracted light curves with a bin size of 32 s 
(standard 2 mode) were extracted in the soft (2 -- 5 keV) and hard energy band (16 -- 30 keV).
An uninterrupted individual segment of long duration was considered and the {\it crosscor} tool implemented in
the XRONOS package was used for the analysis (for more details of the program  {\it crosscor},
see Sriram et al. 2007; Lei et al. 2008; Sriram et al. 2011a).
Generally the shape of cross-correlation function (CCF)                                       
is found to be complex but the peak of CCF can be fitted by an inverted Gaussian function
(Choudhury \& Rao 2004; Sriram et al. 2007; Lei et al. 2008).
We therefore fitted the Gaussian function to the CCF data to calculate delay times and errors
at a 90\% confidence level ($\Delta \chi^{2}$=2.71).
We also used the single bit mode data to analyze the power density spectrum (PDS) of a few observations
discussed later. 

\section{Temporal Analysis Results}
\subsection{Anti-correlated lags in GX 5-1}

We found that most of the observations for GX 5-1 show a positive correlation between the light curves in
2 -- 5 keV and 16 -- 30 keV and a few of them show an ambiguous correlation or a low level of correlation,
i.e. a low correlation coefficient (CC). Those are similar to the observations for Cyg X-2 and BHXBs
(e.g., see Choudhury \& Rao 2004). Details of the 57 observations are given in Table 1. 
It was found that the observations located at the HB and hard apex show a positive correlation, 
a low level of correlation or an anti-correlation without any time lag. 
For example, in ObsID 30042-01-03-00 a clear anti-correlation at $\sim$ --0.52 was observed without lag.
We define the low level of correlation as the CC value of $\le$ $\pm$0.35 and find that many observations fall 
in this category (Table 1). It is also found that during the observations the source was moving/varying toward 
all the possible directions 
along the Z track (i.e. from the upper HB to hard apex and vice-versa). 

The positive correlations are generally expected and can be understood in the simple scenario of
overall intensity variation. 
We, however, detected strong anti-correlated X-ray lags from the light curves of 17 observations. 
The representative samples are displayed in Figure 1.
We carried out the cross-correlation analysis for each segment of the light curve
(we adopted the full light curve in case that the segments showed a similar order of delay for one observation).
The cross-correlation functions (CCFs) as a function of delay are plotted in the right panels of Figure 1.
The segments in which the lags are found are marked by the letters `a' and `b' in the light curves. 
The minima of CCFs, 0.4 -- 0.6, suggest that the anti-correlations are very strong
(the estimated null hypothesis probabilities are $<$10$^{-6}$).
The observations for which the anti-correlated lags were detected in GX 5-1 are listed in Table 2 together
with the delay times that range from a few tens to hundred seconds. 
Among our selected data sets, we found an anti-correlated hard X-ray lag from 13 observations and found an 
anti-correlated soft X-ray lag from 4 observations (Figure 1 and Table 2). 
In one observation, both hard and soft X-ray lags were observed. 
The obtained delay times are similar to those found in Cyg X-2 (Lei et al. 2008) but significantly lower than
those in BHXBs (Choudury \& Rao 2004; Choudhury et al. 2005; Sriram et al. 2009). 

\subsection{Lags and their locations on HID}

We obtained the HID using the observations made in the period of 1997 - 1998 because our
anti-correlated lags were mostly detected during this period.
Furthermore, nine observations were made within a span of two months and hence we presumed that the
HID would not be affected by any long-term secular variations.
The HID shown in Figure 2 clearly displays the Z-shaped track. 
The locations on the HID are listed in Table 2 along with the hardness ratios for each observation.
For the observations beyond the period, i.e., the last four observations in Table 2, we determined
their locations based on the PDS analysis.
We found that majority of the lags are located at the HB and upper NB. There are only a few lags at
the soft apex and FB (see Table 2 and Figure 2). In case of the observations that the lags were detected,
the source remained at the HB and hard apex for $\sim$ 19 hrs, at the NB for $\sim$ 14 hrs, and at the FB and 
soft apex for $\sim$ 8 hrs. 
Moreover, three observations that showed an inconsistent anti-correlated lag remained at the HB and 
upper NB together with a few other observations having a low CC (Table 1).
We also found that out of the 23 detected lags, only 4 lags are negative soft lags which tend to appear near
the hard apex. Figure 3 shows the histograms for the 128 segments of the 74 observations with respect to their 
locations on the Z-track. 
The observations showing smaller CCF than $\pm$0.4 were included in the histograms of anti-correlation data.
It is evident that most of the anti-correlated lags are localized along the HB and hard apex.
  
To investigate the possibility that two uncorrelated light curves give a correlation purely by chance,
we have performed simulations by generating two long independent light curves (2$^{20}$ s) based on the method of Timmer and Koenig (1995).
The observation of ObsID 30042-01-11-00 showed the highest anti-correlation coefficient ($\sim$--0.9) and hence we used 
the parameters of the soft band light curve (i.e., mean count rate of 5229 counts s$^{-1}$, standard deviation of 118 counts s$^{-1}$, and power-law index 
of PDS $\beta$ = 1.76) and hard band light curve (i.e. mean count rate of 191 counts s$^{-1}$, standard deviation of 11 counts s$^{-1}$, and
power-law index of PDS $\beta$ = 2.85) to obtain the two simulated light curves. 
The power-law index was obtained from the observation by fitting a power-law model to the corresponding PDS. Then the two simulated light curves were 
cross-correlated, where the simulated curves have a bin size of 32 s and a segment length of 3000 s. 
The CCF and delay histograms are shown in the Figure 4. The number of segments showing the correlation 
coefficient of $\le$--0.9 are low (22 out of 348 segments).
It was also found that in the simulations the number of segments showing a low level of correlation (or no correlation)
is also very low (see Figure 4).

\section{Spectral Analysis Results}
\subsection{Model-independent study of spectral variation}

There is an idea that the radiative process changes during the lag time.
It is therefore important to confirm whether the spectrum varies during the anti-correlated lag times. 
To see such a change, we extracted energy spectra from the initial and final segments of the light curves
and superposed them. Figure 5 shows the superposed spectra (left panels) along with their spectral ratios
(right panels) for four observations that are located at different branch positions.
The spectral ratios show a spectral pivoting around $\sim$3 -- 10 keV and its energy changes depending on
the location on HID. Similar kind of spectral pivoting was found in BHXBs
(Choudhury et al. 2005; Sriram et al. 2007; Sriram et al. 2009) and Cyg X-2 (Lei et al. 2008).
We also found that for positively correlated observations the spectra do not show such a spectral pivoting, 
similar to the result by Lei et al. (2008). For example, we found no spectral pivoting in the spectra
of ObsID 30042-01-12-00 in which the CCF ($\sim$0.3) shows no correlation or a low level of correlation. 
This model independent analysis suggests that the
underlying radiative process changes during the detected anti-correlated lag times. 
It is interesting to note that the pivoting energy is close to 10 keV at the FB/soft apex and the spectral
ratios show a sharp and discontinuous feature at 10 -- 15 keV. It is tempting
to speculate that fast spectral pivoting at this energy might be the reason for the unexplained
spectral feature detected at 10 keV at the FB (Jackson et al. 2009; Asai et al. 1994).

\subsection{Spectral variation in horizontal branch}

We carried out spectral analysis on the various positions of HB. For this purpose, we selected the data of
ObsID 30042-01-13-00 which spans from the hard apex to upper HB.
In the observation, the soft and hard X-rays were anti-correlated: the decrease in the soft X-rays
accompanied by a simultaneous increase in the hard X-rays (top panel in Figure 6). 
A delay was also observed for each segment of the light curve together with a variation of the spectral pivoting
energy (Table 3), except the segment B which does not show any correlation/anti-correlation (CCF $\le$ 0.2).
We extracted energy spectra for the four segments marked as A, B, C, and D in the light curve of Figure 6 and 
did a model independent comparison for the spectra which strongly suggest a change in the radiative process
(the second panel of Figure 6).
We then fitted two component models to the spectral data of 3 -- 25 keV,
``diskBB +CompTT" (diskBB model: Mitsuda et al. 1984; CompTT model: Titarchuk 1994) 
and ``BB+CompTT", which are often applied to fit 
the spectra of Z-type sources (Done et al. 2002; Agrawal \& Sreekumar 2003; Agrawal \& Misra 2009). 
Due to the absence of data below 3 keV, the absorption column density was fixed at
N$_{H}$ = 6.0$\times$10$^{22}$ cm$^{-2}$ (Jackson et al. 2009). The best-fit spectral parameters are 
summarized in Table 3. 

It is found that as the soft X-ray decreases from the segment A to D, both disk temperature
(kT$_{in}$ = 2.10 keV to kT$_{in}$ = 2.87 keV) and black body temperature 
(kT$_{bb}$ = 1.51 keV to kT$_{bb}$ = 2.04 keV) increases independently.
The electron temperature (kT$_{e}$) does not vary significantly when the BB model is applied 
whereas its temperature considerably increases when the diskBB model is applied,
i.e., as the source moves from the hard apex (segment A) to upper HB (segment D), the electron temperature 
(kT$_{e}$) increases whereas the optical depth ($\tau$) decreases (Table 3).
The unfolded spectra shown in the third  panel from the top of Figure 6 were obtained with the diskBB+CompTT model.
From this spectral analysis, it is inferred that both Comptonizing medium (or Compton cloud) and disk 
properties change when the source moves along the track (from hard apex to upper HB). In case that 
the BB+CompTT model is applied, the black body temperature only varies significantly (Table 3). 

We analyzed PDSs for each of the segments to fully understand the physical changes,
whether there is a dynamical change of the disk or not (i.e. radial movement of the inner disk front;
Done et al. 2007). 
It was found that the four segments (A, B, C, and D) show HBOs having different frequencies (Table 3, the bottom panels of Figure 6), 
which are a general characteristic of Z sources. From the segment A to D, the HBO frequency changes from 
$\nu_{HBO}$ = 51 Hz to $\nu_{HBO}$ = 21 Hz. We found that as the source moves from the hard apex to 
upper HB the HBO frequency decreases, which is in agreement with the result of Jonker et al. (2002) 
who found that low frequency QPOs (15 -- 50 Hz) increase from the upper HB to upper 
NB in GX 5-1. In general, the relatively high frequency QPOs are supposed to originate closer to NS/BH and 
low frequency QPOs originate away from NS/BH in the Keplerian disk. The dynamical variability analysis indicates 
that at the hard apex the inner disk front is close to NS ($\nu_{HBO}$ = 51 Hz) and at the upper HB the disk
is away from NS ($\nu_{HBO}$ = 21 Hz).

In brief, as the source traverses from the
hard apex (section A) to upper HB (section D), the observed delay decreases, HBO frequency decreases, pivoting energy
increases, black body temperature and flux increase, and the Comptonizing region changes from
a low temperature and high optical depth compact plasma to a high temperature and low optical depth plasma. 
If we assume that the Compton cloud resides inside a truncated disk and these changes are
interpreted as an increase in the truncation radius, these findings indicate that there exists
a small black body region possibly at the surface of the neutron star, which is unrelated to the
size of the truncated disk (disk temperature should have considerably reduced, if it
is related to the truncation radius). The delay time, then, should pertain to the 
adjusting timescale for the Compton cloud in the new disk configuration, 
which appears to be greater for the high optical depth plasma.

\subsection{Frequency variation of HBOs}

In the case of BHXBs, it is known that during the lags the QPO frequencies shift to higher or
lower frequencies (Choudhury et al. 2005; Sriram et al. 2007; Sriram et al. 2009; Sriram et al. 2010). 
The origin of HBOs and NBOs in NS-LMXBs are not well known, particularly the generating mechanism of low frequency
QPOs is poorly understood. These QPOs can originate either from the Corona
(Chakrabarti \& Manickam 2000; Titarchuk \& Fiorito 2004) or from the disk (Done et al. 2007; Ingram \& Done 2010). 
To see the high frequency variation of HBOs in detail, we select ObsID 20053-02-01-01 which shows a negative
lag of $\sim$ --250 s, because Wijnands et al. (1998) already reported the kHz QPO of $\sim$500 Hz for this 
observation.

We analyzed the PDS for ObsID 20053-02-01-01 shown in Figure 7 (which is obtained from the initial (A) and final (B) 
segment of the light curve in Figure 1) and found that there is a significant frequency variation in the HBOs 
from $\nu_{HBO}$ = 23 Hz (segment A) to $\nu_{HBO}$ = 18 Hz (segment B), suggesting a dynamical change in the
accretion disk.
Wijnands et al. (1998) found that there are kHz QPOs in the higher energy band (8.6 -- 60.0 keV)
but not in the lower energies.
For comparison, we extracted the Leahy power PDS for the segment A and B in high (8.6 -- 60.0 keV) and 
low (6.9 -- 8.3 keV) energies and plotted those in Figure 7.
As shown in Figure 7 and Table 4, there exist kHz QPOs at $\sim$590 Hz (segment A) and $\sim$550 Hz (segment B)
in the higher energy band only, confirming the earlier result by Wijnands et al. (1998). 
Apart from the frequency variation in HBOs, we found that the frequency in kHz QPOs tends to decrease from
$\sim$590 Hz to $\sim$550 Hz although this tendency is marginally significant. 
      
We also investigated a spectral variation between the two segments.
For this, we applied the diskBB+CompTT model to the spectra and extracted the unfolded spectra
shown in Figure 7. The best-fit parameters are listed in Table 4.
By applying the best-fit parameters of the spectrum A to B, we obtained the $\chi^{2}$-value of
$\chi^{2}$/dof = 3027/47 which clearly indicates that the spectrum has changed.
We found that the disk and thermal Comptoinzation parameters have changed significantly. For example, 
the disk temperature has increased from kT$_{in}$ = 2.84 keV to kT$_{in}$ = 2.95 keV along with the increase of
electron temperature from kT$_{e}$ = 7 keV to kT$_{e}$ = 13 keV.
Similar kind of variation in electron temperature was reported for GRS 1915+105 in which a lag was detected 
(Sriram et al. 2007). The changes in HBO (from 18 Hz to 23 Hz) and kHz QPO (590 Hz - 550 Hz) along with the 
electron temperature (7--13 keV) clearly indicates that the Compton cloud size is increasing which can be the cause of 
the observed soft X-ray lag in this observation.

\section{Discussion and Conclusion}
\subsection{Anti-correlated lags and their positions on HID}

We investigated the CCFs of GX 5-1 between soft (2 -- 5 keV) and hard (16 -- 30 keV) X-ray light curves.
Lei et al. (2008) detected anti-correlated hard and soft lags for Cyg X-2 and they suggested
that the accretion disk in the source could be truncated. O' Brien et al. (2004) also found
an anti-correlated variation for Cyg X-2 between optical and X-ray. 
Earlier studies of a few BHXBs reported a few 100 -- 1000 s lags in the CCF of soft and hard X-ray bands 
and the detailed temporal and spectral analysis favored a truncated 
accretion disk geometry too (Choudhury et al. 2005; Sriram et al. 2007; Sriram et al. 2009; 
Sriram et al. 2010). The truncated accretion disk model is often used to explain the spectral and temporal 
features of the hard states observed in BHXBs  (Done et al. 2007) but some observational results show that 
the disk may not
be truncated (Miller et al. 2006; Rykoff et al. 2007). Done \& Gierlinski (2006) argued that non-truncated
disk models can be ruled out based on the outflowing disk wind processes. The broad iron line in low/hard
state was presented as the evidence of non-truncated disk (Miller et al. 2006) but a re-analysis of the
data showed that the line is instead narrow and consistent with the truncated disk geometry 
(Done \& Diaz Trigo 2010). The physical condition is relatively more complex in neutron star LMXBs. 
We carried out such a similar study for GX 5-1 because the longer time lags seem to provide an opportunity
to explore the geometry of accretion disk in LMXBs.

Our study on CCFs showed that on most occasions the soft and hard X-ray light curves are positively
or loosely correlated but on a few occasions the CCFs show statistically significant anti-correlated 
hard (mostly) and soft X-ray lags (Figure 1, Table 1, and Table 2).   
Since the hard X-rays are Comptonized from the 
soft X-rays, they should be correlated each other but this and previous studies for a few BHXBs and Cyg X-2
indicate that a re-adjustment in the emission geometry and radiation process is necessary to interpret the 
anti-correlations.
The hard X-ray lag means that the arrivals of hard X-rays are delayed to the observer compared with the soft
X-rays. We found that the centroid frequency in HBOs changes toward the lower frequency during the lag (Figure 6). 
This result strongly suggests that the accretion disk dynamically changes during the lag. 
The frequency shift could be due either to an outward radial movement of disk 
(Done et al. 2007) or to change in the oscillation of Compton cloud (Chakrabarti \& Manickam 2000). 
The difference in lag timescale implies that the radial movement of disk occurs differently at various radii/locations 
which can effectively affect on the properties of Compton cloud. If we assume that the mass accretion rate 
is the primary parameter to determine the location on HID,
then the detected lags should be related somehow to the viscous timescale during which the accretion
disk is re-adjusted because the other timescales like dynamical and thermal timescales in accretion disk are 
less in magnitude (Frank et al. 2002). At the same time the observed lags are much smaller than the 
viscous timescale of the full accretion disk, which is generally of
the order of a few tens to hundred days (often assumed to be the duration of an outburst) (Sturner \& Shrader 2005 ),
and hence we speculate that the inner disk front movement occurs at a much smaller radius. 
However, based on recent results (Lin et al. 2009; Homan et al. 2010), mass accretion rate was 
ruled out as the primary driving physical parameter to cause a Z shape in HID hence we suggest that the observed 
lags could be due to change in the size of the Comptonizing region. Based on the study of black hole source 
XTE J1550-564, Homan et al. (2001) found that independent of mass accretion rate, the size of the Comptoinzing region 
could be potentially possible physical factor to determine the state of the source.  

Our observed lags may not be explained by invoking just a simple radiative process such as Comptonization. 
There are several models which can account for small ($\le$ 1 s) timescale lags (Kazanas et al. 1997;
Nowak et al. 1999; Bottcher \& Liang 1999). In these models, some of the soft X-rays radiating from the
NS surface, accretion disk, or a soft X-ray source in the form of a dense blob made of disk material 
get Comptonized by hot electrons ($\sim$10$^{8}$ K) and are converted into the hard X-rays.
During this process the hard X-rays can be delayed by $\le$ 1 s over the soft X-rays that do not suffer 
from the Comptonization. However, we detected about hundred times longer lags from  GX 5-1.
These lags can be explained if we assume that the accretion disk is truncated.
Jackson et al. (2009) carried out a detailed spectral analysis for GX 5-1 and they suggested that
the existence of a truncated accretion disk structure cannot be ruled out at the upper NB and HB.

Similar to Cyg X-2, we found that most of the anti-correlated lag observations are located at the HB and upper NB 
(Table 2, Figure 2 and Figure 3). As such there is no exact relation between observed delays and corresponding positions 
on HID but we suggest that especially at the HB and upper NB, the disk forms a thin structure and corona 
forms an inner quasi-spherical shape making a truncated accretion disk geometry. Fender (1999) found that for 
black hole binaries the physical body forming the Comptonizing corona is the base of the jet.  
The radio observations show that a jet component is possibly present close to the central region of the disk 
when the source is at the hard apex and HB (Penninx 1989; Berendsen et al. 2000). Hence 
we suggest that the detected lags indirectly indicate a truncated accretion
disk in GX 5-1 when the source is at the HB and NB. Since the mass of NS in GX 5-1 is not known, it is
difficult to estimate the truncation radii. If we assume the properties of Cyg X-2,
the truncation radius would be slightly less than 18 NS radius (Lei et al. 2008). It is because,
GX 5-1 is fainter than Cyg X-2 and the viscous timescale is proportional to the inverse square of mass 
accretion rate (t$_{vis}$ $\propto$ ${\dot M}$$^{-2}$ ).

\subsection{Spectral variations}

The model independent spectral analysis in \S4.1 suggests that the radiative process changes during the
lag time which can be clearly seen in the spectral ratios in Figure 5. 
The spectral ratios show that the intensity variation accompanies a spectral hardening
in three observations but accompanies a spectral softening in the other observation.
This spectral pivoting is a strong signature of an opposite radiative process change in the
soft and hard X-ray emitting regions. Such a spectral pivoting was not observed when the source shows 
a correlated intensity variation. Lei et al. (2008) reported a similar result too.
  
We analyzed a spectral and PDS evolution from  the hard apex to upper HB (segment A to D)
and obtained that as the HBO frequency decreases ($\nu$ = 51 Hz at the hard apex to $\nu$ = 21 Hz at
the upper HB), the coronal temperature increases along with a decrease in the optical depth (Table 2).
Simultaneously, the increase in the disk temperature is accompanied by a decrease in the 
disk normalization which is a close measure of the inner disk radius. 
In addition, the variation in the HBO frequency is consistent with the variation of coronal parameters 
(Chakrabarti \& Manickam 2000; Titarchuk and Shaposhnikov 2005) but is
inconsistent with the variation of disk or black body parameters.
This discrepancy also arises when we relate the HBO frequency variation with that of diskBB/BB spectral
parameters, i.e., in the frame of a truncated accretion geometry, 
as the disk moves inward (in a dynamical point of view, $\nu_{HBO}$ = 21 Hz
to 51 Hz), the disk temperature is supposed to increase but it is decreasing. It was found that in a strong 
Comptonization the obtained disk temperature should be lowered (Done \& Kubota 2004).

Conventionally as the disk approaches close to the NS, the QPO frequency increases, 
disk normalization decreases and the increase in soft
X-rays cools the relatively hot corona. Since the disk is in a deeper gravitational potential, the inner disk
temperature should be relatively high when compared to the disk temperature at lower QPO frequency.
But our analysis shows that the disk temperature is low along with a high disk normalization value 
at higher HBO frequency ($\sim$51 Hz)  whereas at low HBO frequency ($\sim$21 Hz) the disk 
temperature is high with a low normalization value. Even the black body temperature is found to be low at 51 Hz compared with the BB 
temperature at 21 Hz.
This decrease in the soft component temperature is similar to the result by Jackson et al. (2009)
and Balucinska-Church et al. (2010) who found that, for GX 5-1 and other Z-track sources, the black body
temperature (BB) decreases as the respective source moves from the upper HB to hard apex. 
We suggest that the low inner disk temperature and high disk normalization at 51 Hz HBO (which is implying the disk 
is truncated at large radii) may be due to an instability caused by the radiation pressure ($\propto$ T$^{4}$)
(Jackson et al. 2009). But in this scenario, if the inner region of the disk outflows, the 51 Hz HBO should not be observed. 
The other possible scenario to account for the 51 Hz HBO could be that some portion of the coronal material 
condenses back to form the disk in the inner region and which is not in thermal equilibrium.
Such type of re-condensation of corona in the intermediate/SPL state is explored in black hole
binaries (Meyer et al. 2007) and the observational signatures were found in GX 339-4 (Sriram et al. 2010). 
However, such re-condensation models are yet to be explored for neutron star sources.   
Alternatively, the observed black body component may not be arising from the inner parts of the truncated disk;
rather they may be a part of the NS surface.

We also carried out the spectral and PDS analysis for ObsID 20053-02-01-01 in which a soft X-ray lag
($\sim$ -250 s) was detected.
It was found that the HBO frequency shifts from $\nu$ = 23 Hz (segment A) to $\nu$ = 18 Hz (segment B)
whereas the disk temperature increases from  kT$_{in}$ = 2.84 keV to kT$_{in}$ = 2.95 keV
with a simultaneous decrease in the disk normalization (Table 4). 
During these change, the electron temperature increases from kT$_{e}$ = 7 keV to kT$_{e}$ = 13 keV 
with a decrease in the optical depth. 
The frequency shift suggests that the disk is moving away from the NS along with an increase in the electron 
temperature, which is generally observed in BHXBs. But the disk temperature increases instead of  
decreasing. Again these changes can be explained in the frame work of a re-condensation process. 
The radial outward movement of disk (based on the HBO shift) and the expansion of Comptoinizing region 
could cause the soft X-ray lag. Li et al. (2007) explored such kind of disk movement theoretically and 
found that fluctuations in the inner region of the disk can disturb the given configuration
and therefore this disturbance can cause the disk movement with a timescale of a few hundred seconds.

\subsection{Truncated disk geometry: Overall analogies of lagged observations in BHXBs and Z sources}

In case of BHXBs, a detailed spectral analysis presented that the following parameters, e.g., the physical 
size of accretion disk, the electron temperature (kT$_{e}$), and the optical depth of Compton cloud,
are all changing for the observations that lags are detected.
It was suggested that as the inner front of accretion disk moves toward the black hole, 
the soft flux increases and cools the hot electrons which results in the relative decrease of hard X-ray flux
(Sriram et al. 2007; Sriram et al. 2009; Sriram et al. 2010).
The lag time is comparable to the viscous timescale during which the various physical and radiative
process change. 
A recent study reported that the variation in disk emission leads to the hard X-ray emission of a time delay of
$\le$ 1 s or a few seconds, which may be caused by the fluctuations in the disk (Uttley et al. 2011).
In our present work, both disk and Compton cloud properties are changing during the lag time of 
a few tens of seconds. 
In BHXBs, however, the soft X-rays originate from the disk whereas in NS systems the emitting site 
could be either NS surface or accretion disk.

The anti-correlated hard X-ray lags in BHXBs were detected when they are
in the steep power-law (SPL) or intermediate states (for a review of the spectral states of BHXBs see e.g., 
Remillard \& McClintock 2006; Done et al. 2007; Belloni 2010). The intermediate frequency QPOs ($\ge$ 1 Hz) and
kHz QPOs in BHXBs were detected only in these states and radio studies showed that the jet component switches on/off
during these states (Fender et al. 2009). 
A detailed radio study provided a substantial observational evidence for the similarity between Z-type
sources and transient black holes at very high accretion rates (Migliari \& Fender 2006).
Moreover, the Z-type source luminosities close to the Eddington limit (L $\sim$ L$_{Edd}$) which
is often observed in the SPL state of BHXBs (Done et al. 2007). 
The temporal and spectral properties in the SPL/IM state together with the properties in radio emissions
are analogous to those for Z-type sources in the HB and upper NB. One of the important spectral properties 
in the SPL state is the presence of compact corona (Done \& Kubota 2006; Sriram et al. 2007). 
Such signatures for Z-type sources were also revealed by the {\it Integral} observations
(Paizis et al. 2005, 2006). These analogies in Z-type sources and BHXBs suggest that
the truncated accretion disk is the most probable physical configuration which can consistently explain
the various spectro-temporal properties.

In conclusion, we detected anti-correlated hard and soft time lags of the order of a few tens to hundred seconds
in the Z-type source GX 5-1. We found that most of the lagged observations are located at the HB and upper NB.
The magnitude of the lag corresponds to the viscous timescale of the inner accretion disk and hence 
indicates the viscous re-adjustment timescale of a hypothetical truncation radius. The change in QPO frequency
and spectral parameters also support this conclusion.
On the different branches, the spectral and temporal behaviors suggest that during the observations
the accretion disk is dynamically changing, most probably due to the different mass accretion rate.
However we do not find any tight correlation between delay times and branch positions on HID.
The PDS analysis shows that the shifts in HBO frequencies are strongly correlated with the thermal
Comptonization parameters instead of the disk parameters, which suggests that the HBOs are occurring
at the hard X-ray emission region. The spectral
and HBO frequency changes favor the scenario where the inner disk is truncated due to the radiation pressure at the HB and hard apex.
The observed lag and spectral changes made it substantially evident that the geometry of the accretion disk is truncated.
Overall, from the upper HB to hard apex, the spectral and PDS evolution suggest that there is a slightly 
different physical/geometrical condition compared to BHXBs.
In BHXBs, as the QPO frequency decreases, the disk temperature decreases but the coronal
temperature increases. On the other hand, in GX 5-1, both the disk and coronal temperatures increase
as the QPO frequency decreases.
This difference is most probably due to the existence of a hard surface in neutron stars which
can provide a plausible site for the thermal emission.

\acknowledgements
We thank the anonymous referee for the very useful comments.
This research has made use of data obtained through the HEASARC Online Service, provided by the NASA/GSFC, 
in support of NASA High Energy Astrophysics Programs.
\clearpage

\begin{figure}
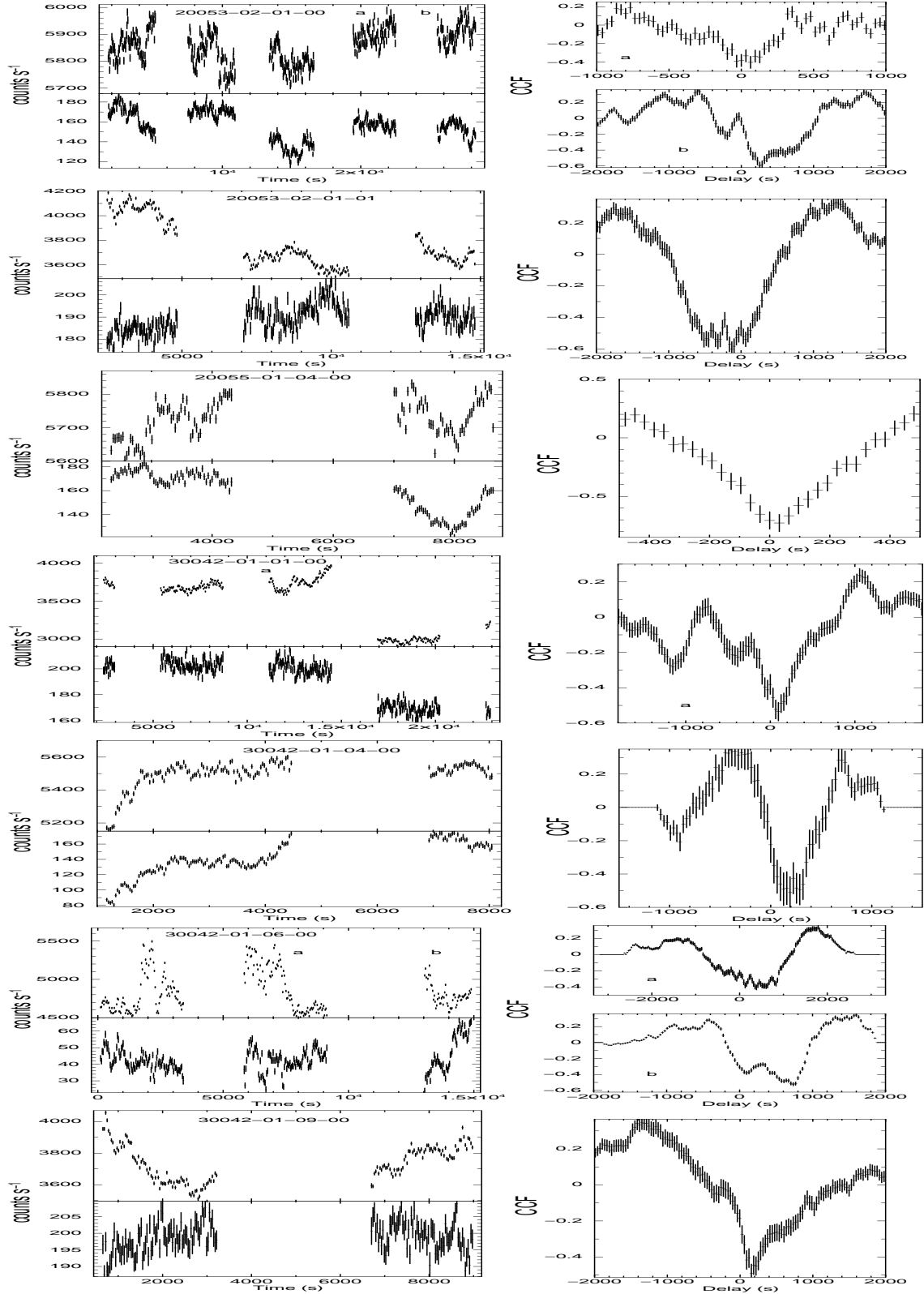

\begin{center}
\includegraphics[width=3.0cm,height=15.1cm,angle=270,clip]{fig1.1.ps}\\
\includegraphics[width=3.0cm,height=15.1cm,angle=270,clip]{fig1.2.ps}\\
\includegraphics[width=3.0cm,height=15.1cm,angle=270,clip]{fig1.3.ps}\\
\includegraphics[width=3.0cm,height=15.1cm,angle=270,clip]{fig1.4.ps}\\
\includegraphics[width=3.0cm,height=15.1cm,angle=270,clip]{fig1.5.ps}\\
\includegraphics[width=3.0cm,height=15.1cm,angle=270,clip]{fig1.6.ps}\\
\includegraphics[width=3.0cm,height=15.1cm,angle=270,clip]{fig1.7.ps}\\
\end{center}
\caption{\scriptsize{The background subtracted light curves of 17 observations for which anti-correlated X-ray
lags are detected (left panels; top: 2 -- 5 keV and bottom: 16 -- 30 keV with bin size of 32 s) and 
the corresponding cross-correlation functions (right panels). The ObsId is labeled for each observation. 
Symbols `a' and `b' correspond to the individual segments for which delays are detected.}}
\end{figure}

\clearpage
\pagestyle{empty}
\begin{center}
\vspace*{-27mm}

\includegraphics[width=3.0cm,height=15.1cm,angle=270,clip]{fig1.8.ps}\\
\includegraphics[width=3.0cm,height=15.1cm,angle=270,clip]{fig1.9.ps}\\
\includegraphics[width=3.0cm,height=15.1cm,angle=270,clip]{fig1.10.ps}\\
\includegraphics[width=3.0cm,height=15.1cm,angle=270,clip]{fig1.11.ps}\\
\includegraphics[width=3.0cm,height=15.1cm,angle=270,clip]{fig1.12.ps}\\
\includegraphics[width=3.0cm,height=15.1cm,angle=270,clip]{fig1.13.ps}\\
\includegraphics[width=3.0cm,height=15.1cm,angle=270,clip]{fig1.14.ps}\\
\includegraphics[width=3.0cm,height=15.1cm,angle=270,clip]{fig1.15.ps}\\

%\centerline{Fig. 1. --- Continued}
\end{center}
\vspace*{17mm}
%\clearpage
\pagestyle{empty}
\begin{center}
\vspace*{-27mm}
\includegraphics[width=3.0cm,height=15.1cm,angle=270,clip]{fig1.16.ps}\\
\includegraphics[width=3.0cm,height=15.1cm,angle=270,clip]{fig1.17.ps}
%\centerline{Fig. 1. --- Continued}
\end{center}
\clearpage

%\clearpage
%\newpage

\begin{figure}
\includegraphics[height=10cm,width=10cm, angle=0]{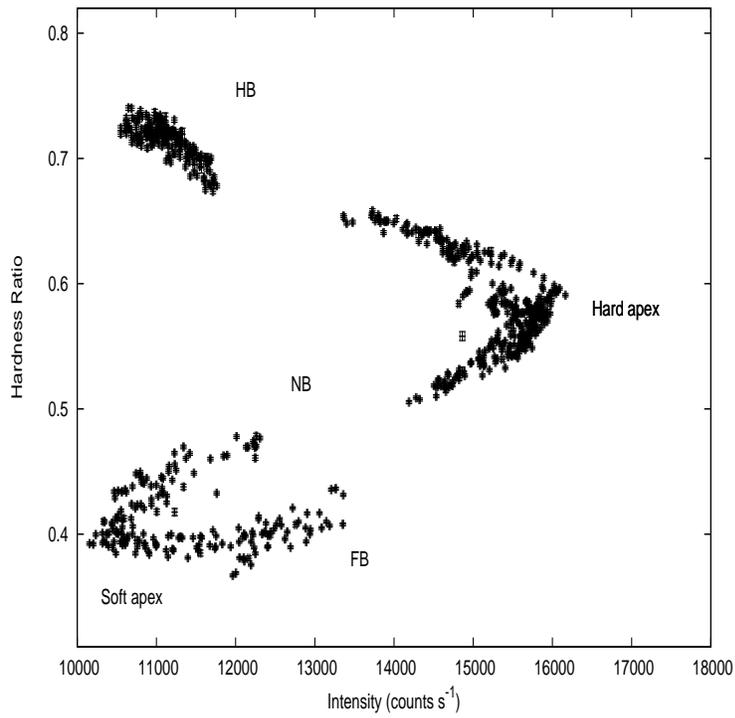}\\

\caption{The Z-track is obtained using the observations of the period 1997-1998 for which 
anti-correlated lags are detected (see Table 2 for the position of each observation).} 
\end{figure}

\clearpage

\begin{figure}
\includegraphics[height=15cm,width=15cm, angle=-90]{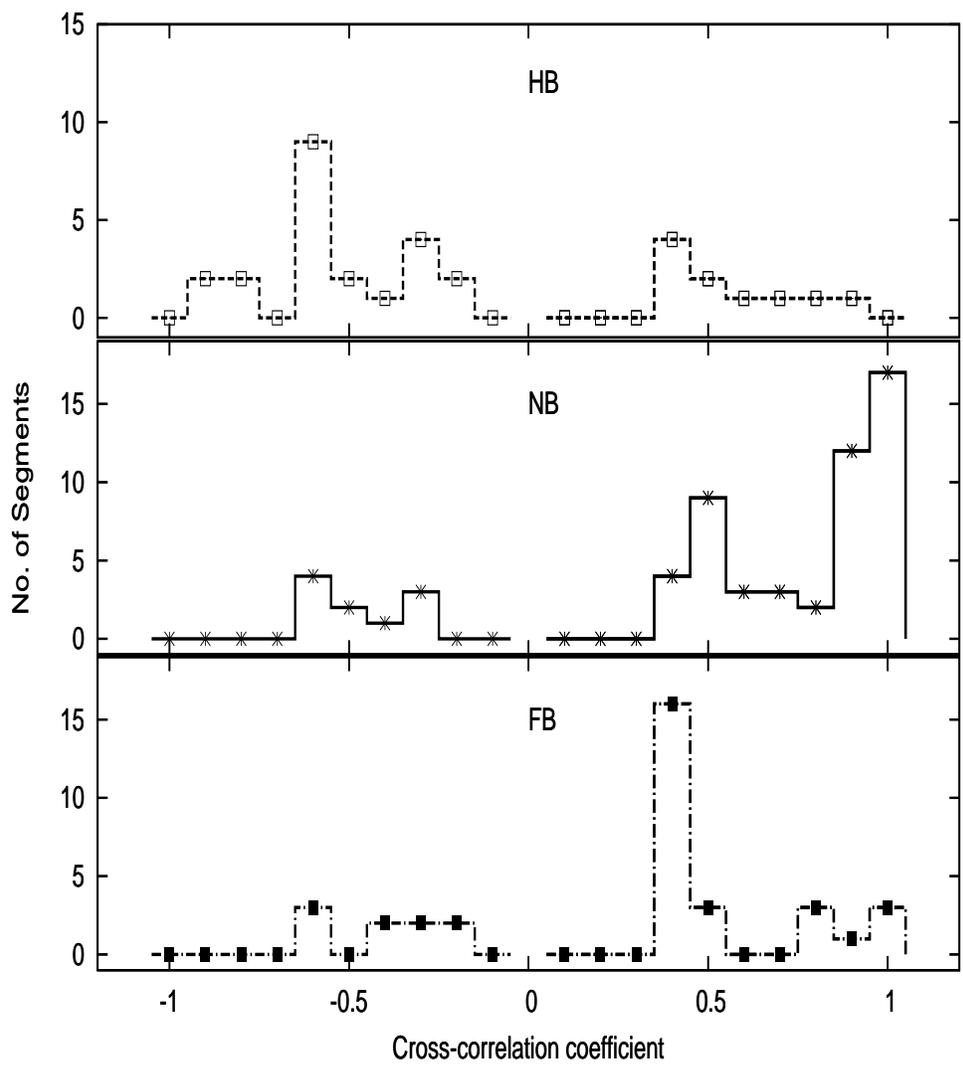}\\

\caption{The histograms of all the observations based on their corresponding locations on the Z-track.
 Hard apex observations were included in HB and soft apex observations were included in FB to obtain the histogram.
 It is clear that most of the anti-correlated observations belongs to HB.} 
\end{figure}

\clearpage
\begin{figure}
\includegraphics[height=17cm,width=8cm, angle=-90]{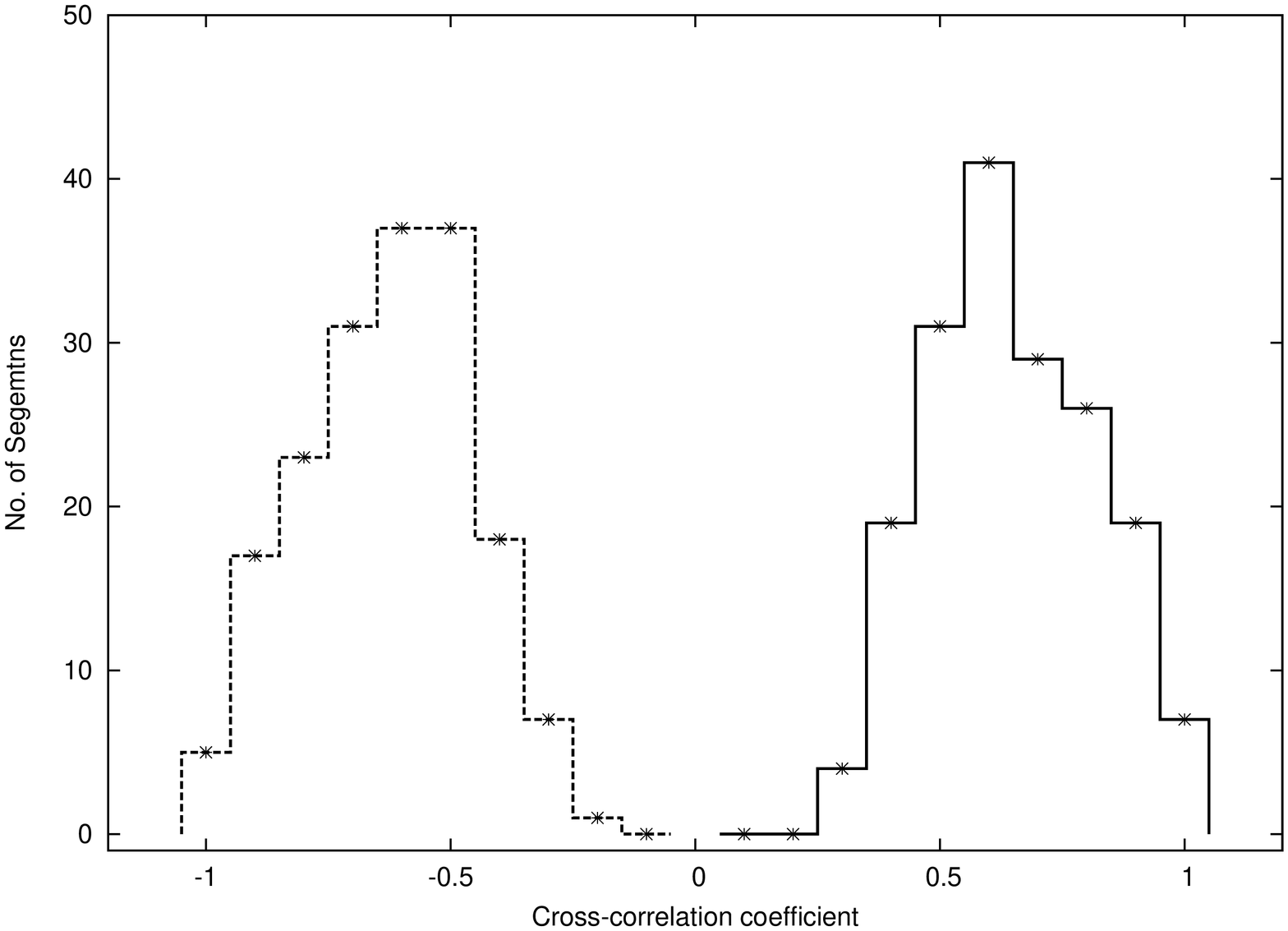}\\
\includegraphics[height=17cm,width=8cm, angle=-90]{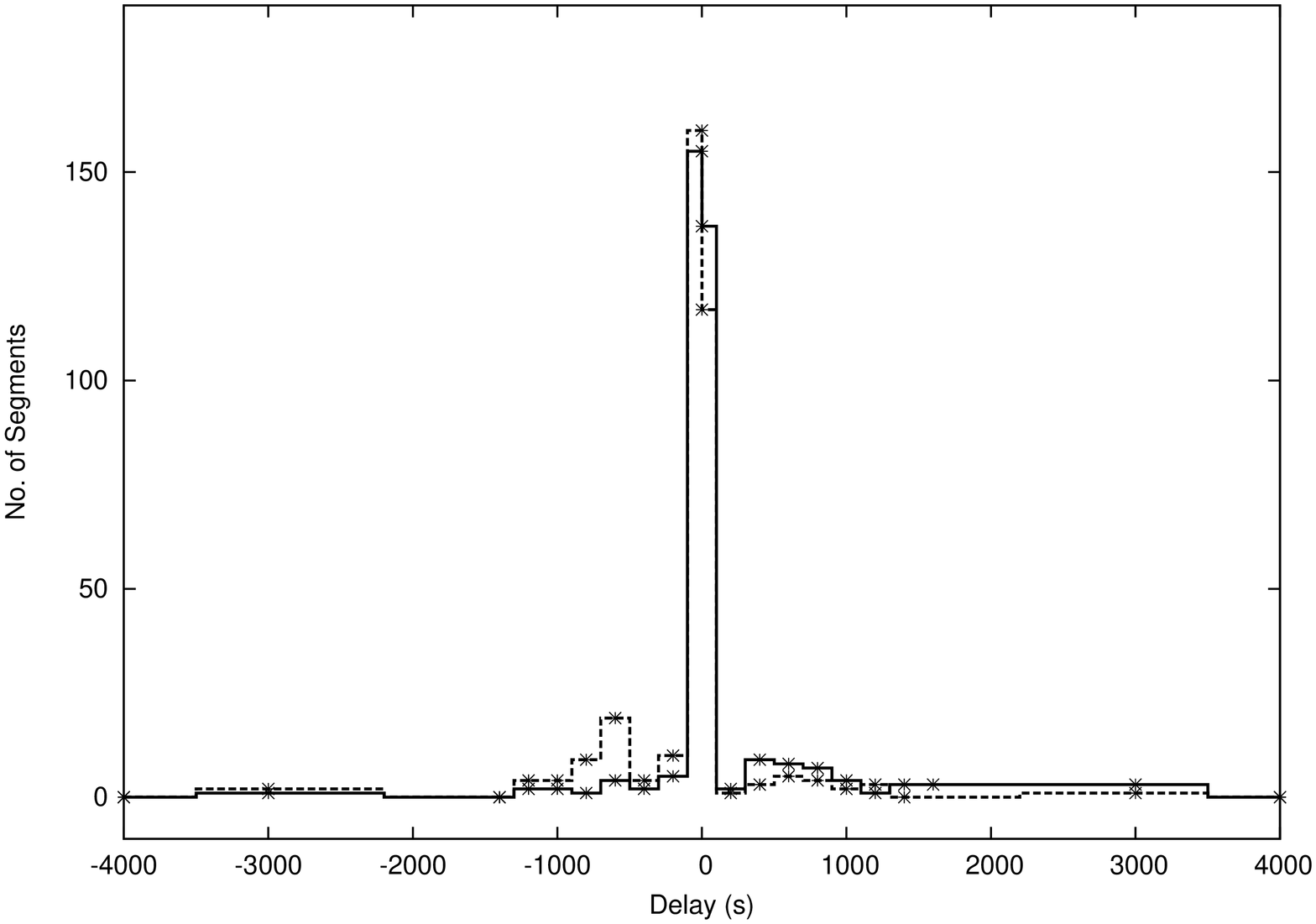}
\caption{ \scriptsize{The upper panel shows the histogram of the CCF obtained from simulated light curves. 
The thick line shows the cross-correlation coefficient when the cross-correlation function is correlated and dashed line shows
the same when the cross-correlation function is anti-correlated. The lower panel shows the 
histogram of the delay, where the thick line shows the delay when the cross-correlation function 
is correlated and dashed line shows the delay when the cross-correlation is anti-correlated (see text).}}

\end{figure}

\clearpage

\begin{figure}
\includegraphics[height=5cm,width=10cm, clip ]{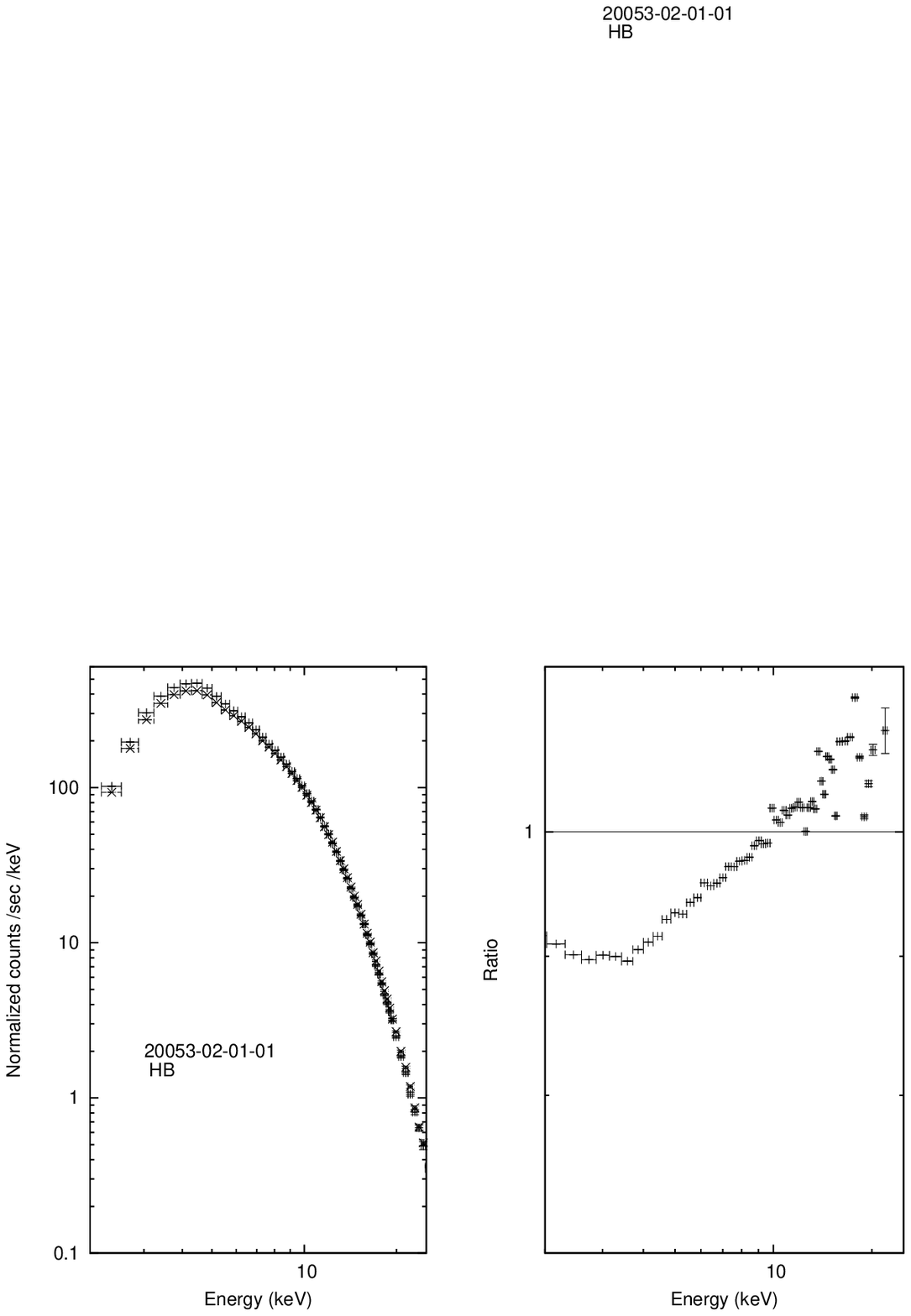}\\
\includegraphics[height=5cm,width=10cm, clip]{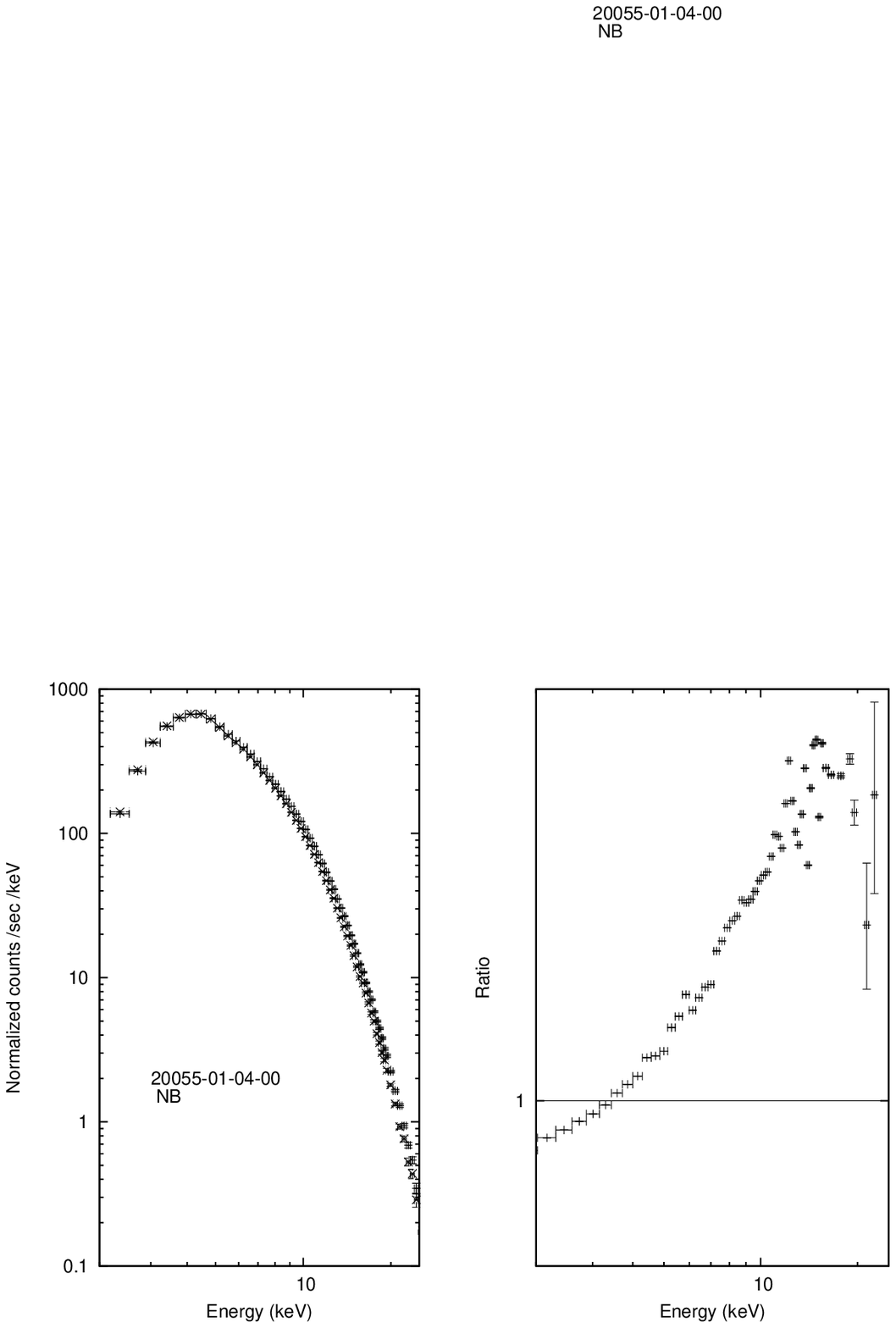}\\
\includegraphics[height=5cm,width=10cm, clip]{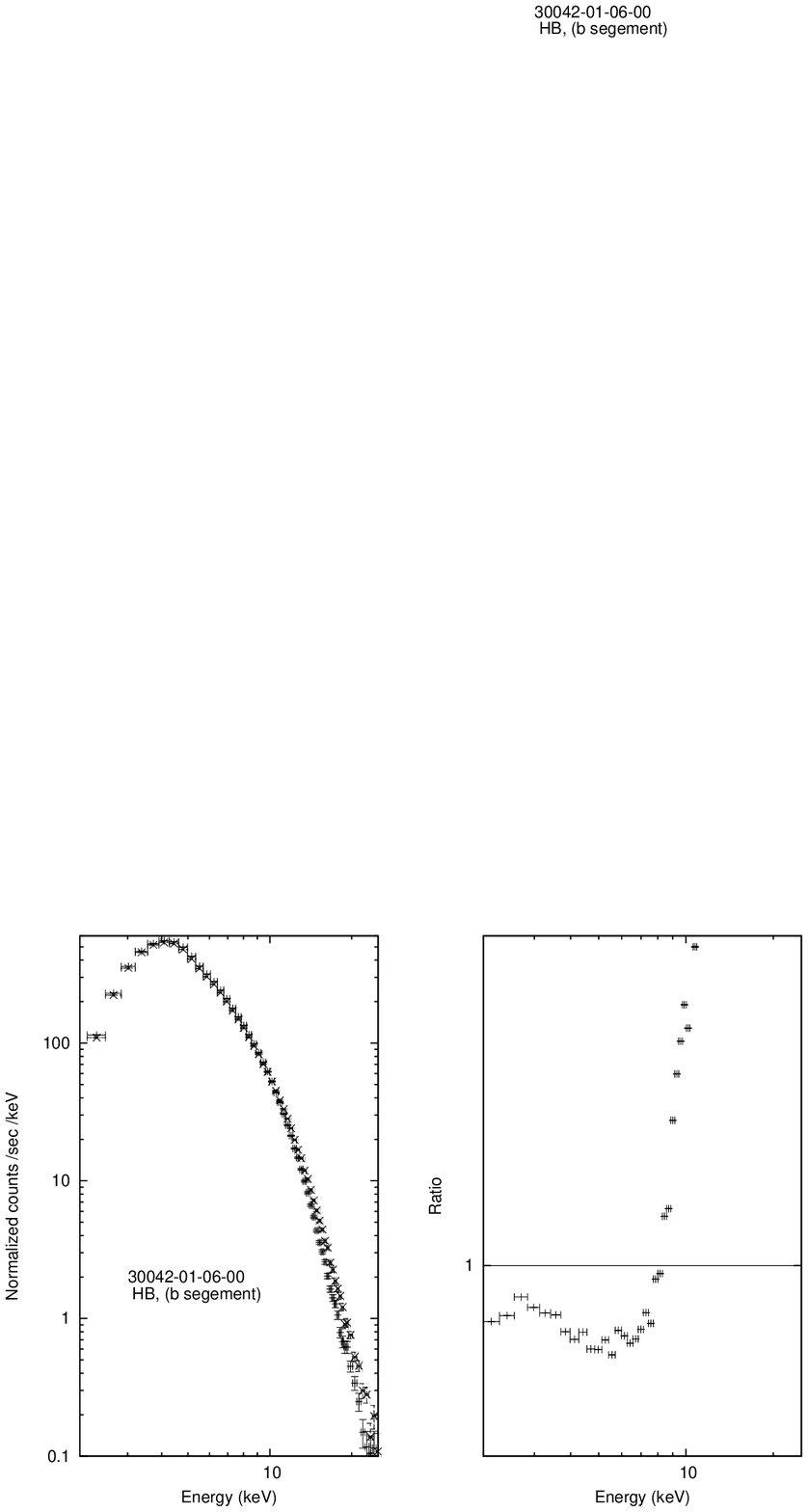}\\
\includegraphics[height=5cm,width=10cm, clip]{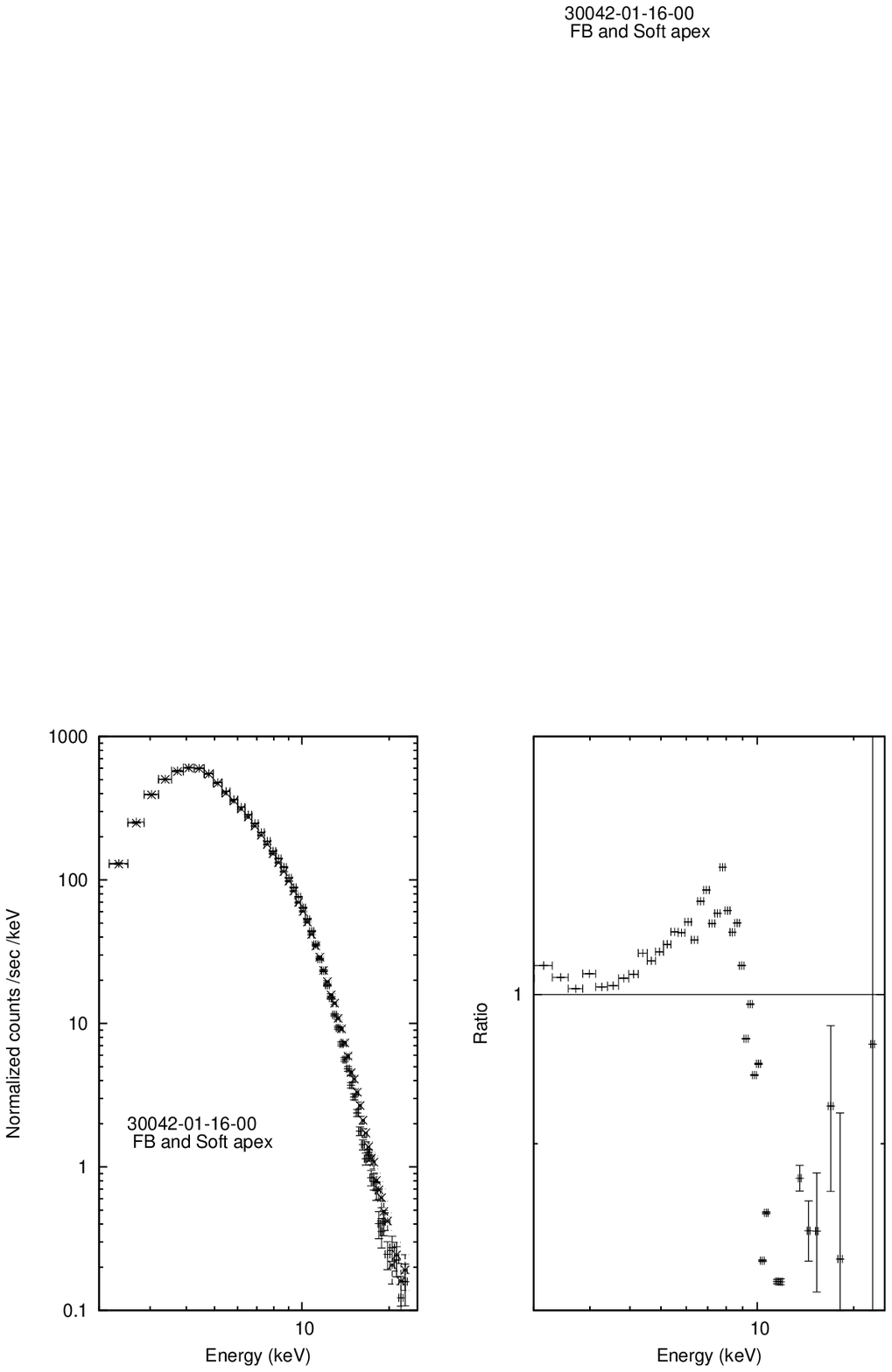}\\

\caption{The energy spectra (left panels) along with their ratios (right panels) are shown for the 
observations that belong to different locations on the Z-track. The ObsID and the position
on the Z-track are given in each figure of the spectra.}

\end{figure}

\clearpage

\begin{figure}

\includegraphics[width=4.5cm, height=17.0cm,angle=270]{fig_6.1.ps} \\
\includegraphics[width=4.5cm, height=17.0cm,angle=270]{fig_6.2.eps}\\
\includegraphics[width=5.0cm, height=17.0cm,angle=270]{fig_6.3.eps}\\
\includegraphics[width=6.0cm, height=17.0cm,angle=270]{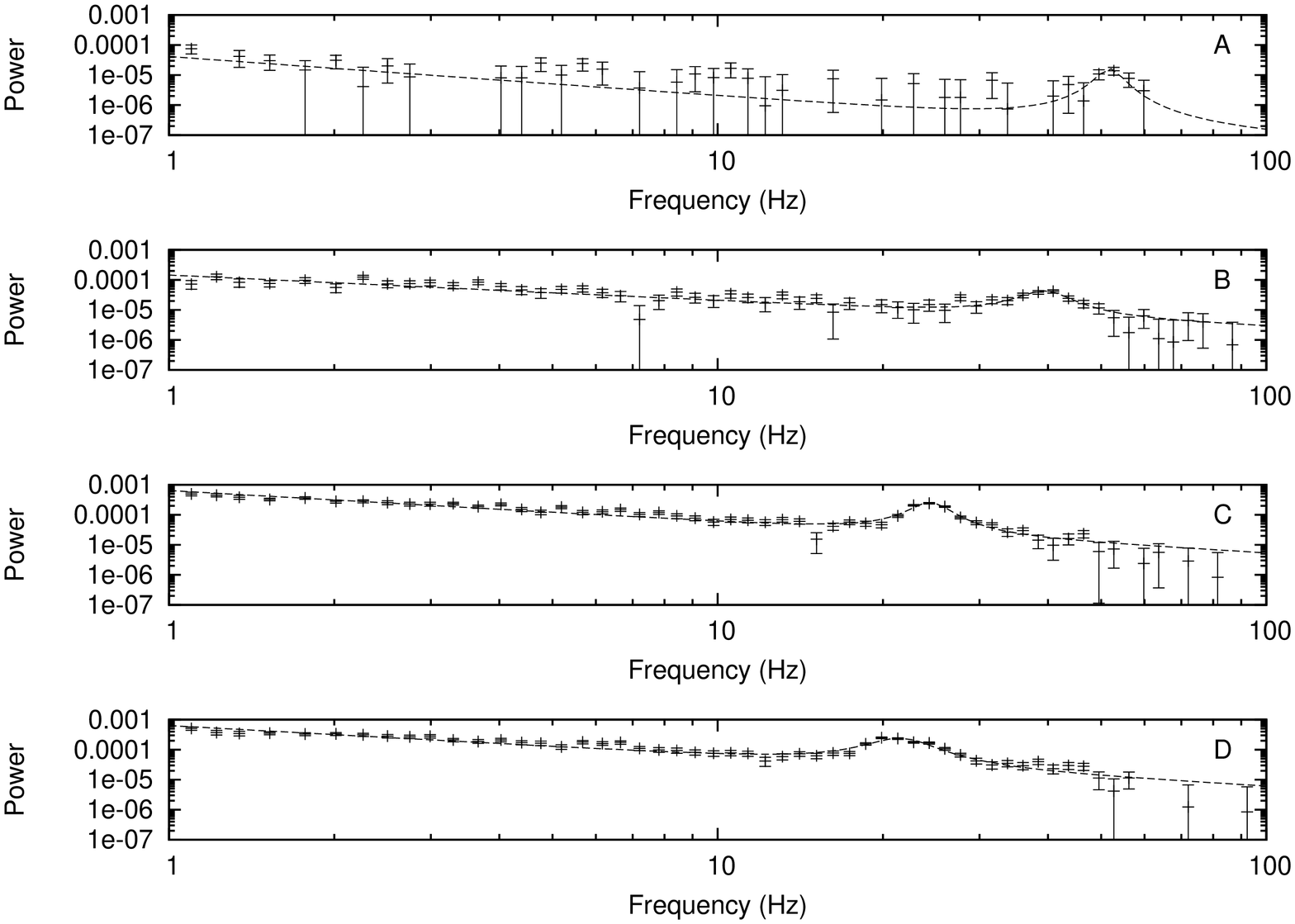}

\caption{\scriptsize{Top: The light curves of soft and hard X-rays for ObsID 30042-01-13-00. 
Second: Model independent spectral variation for the segments (A, B, C, and 
D) is shown. 
Third: The unfolded spectrum for each segment is plotted along with its model components (diskBB+CompTT) 
and the right panels show the $\Delta \chi$ of the corresponding best fit. 
Bottom: Power density spectrum (2 -- 5 keV) for each segment is shown, where 
the power is the normalized power in units of (rms/mean)$^{2}$/Hz. 
The dashed lines represent the best-fit model, power-law + Lorentzian.}}
\end{figure}

\clearpage

\begin{figure}
\includegraphics[height=17cm,width=4.5cm, angle=270,clip=]{fig7.1.ps}\\
\includegraphics[height=17cm,width=5cm, angle=270,clip=]{fig7.2.ps}\\
\includegraphics[height=17cm,width=5cm, angle=270,clip=]{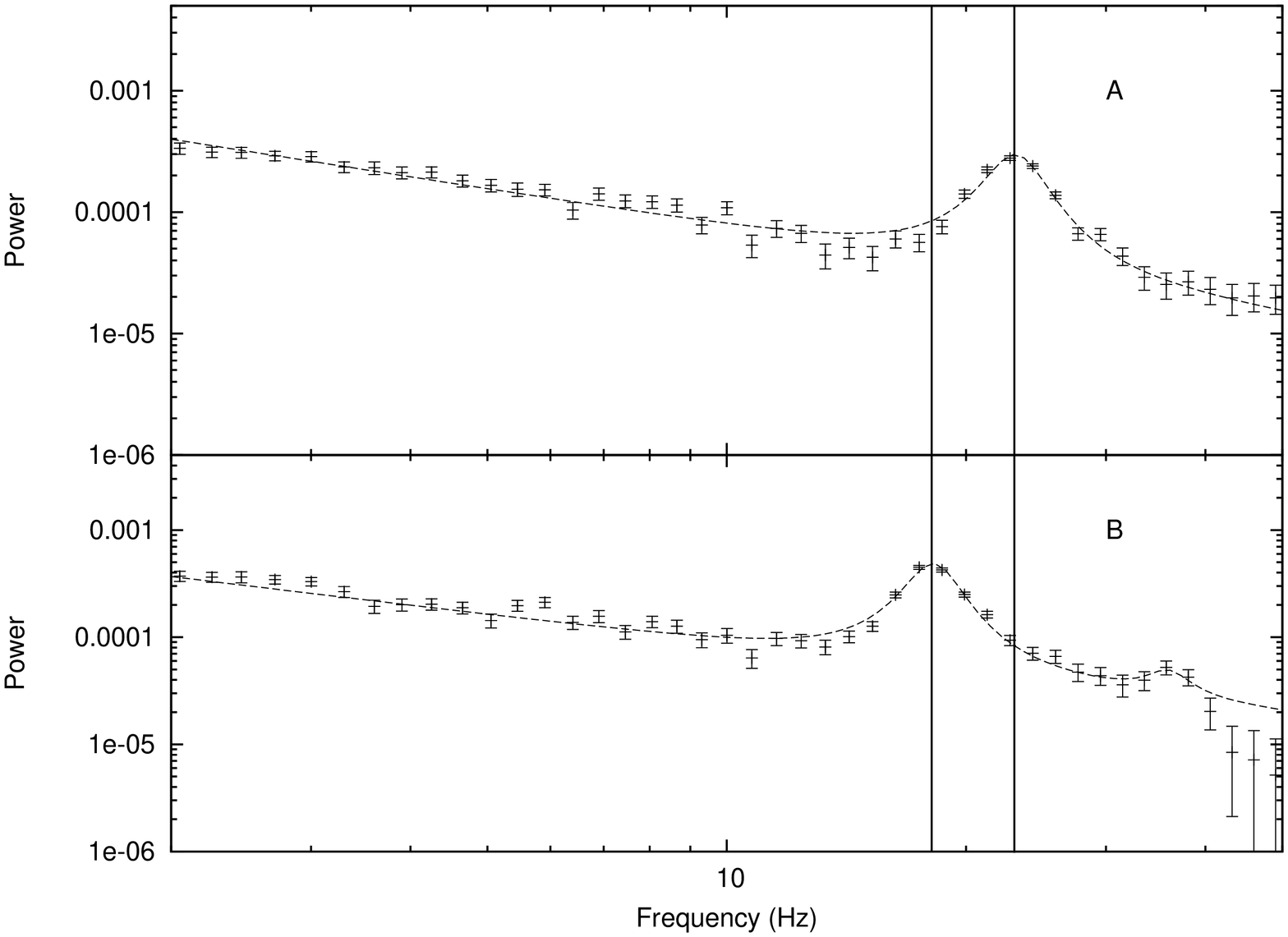}\\
\includegraphics[height=17cm,width=5cm, angle=270,clip=]{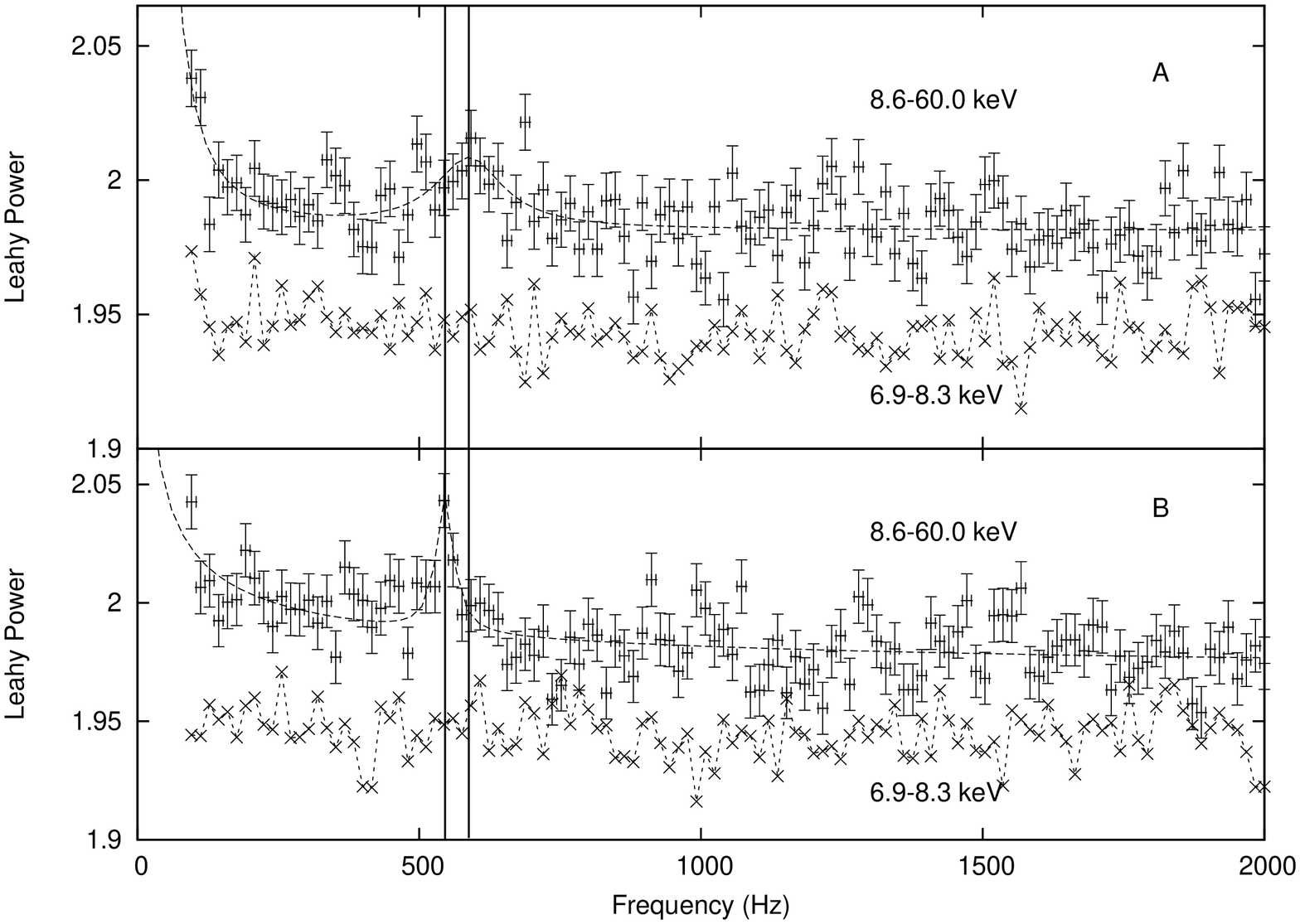}\\

\caption{\scriptsize{Top: Unfolded spectrum of the segment A and B for the ObsID 20053-02-01-01. 
Second panel shows the model independent change between A and B spectra. 
Third panel shows the PDS for the same segments, where the power is in units of (rms/mean)$^{2}$/Hz and 
the dashed lines represent the best-fit model, power-law + Lorentzian.
The vertical solid lines represent the centroid frequencies of the individual segments. 
Bottom: Leahy power PDSs for the segment A and B in the energy band of 8.6 -- 60 keV and 6.9 -- 8.3 keV. 
It is clear that the kHz QPOs are not present in the lower energy band.}}
\end{figure}
\clearpage

\begin{table}
%\fontsize{1pt}
\tiny
\begin{minipage}[t]{\columnwidth}
\caption{\scriptsize{Log of observations showing a positive correlation, a low  level of correlation or an 
anti-correlation without any time lags. Hardness ratios and their branch positions on the Z-track are
	also listed.
For long observations, the range of correlation coefficient (CC) are mentioned if the variations of CC are
observed. The errors in CC (standard deviation) are ranging from $\pm$0.01 -- $\pm$0.05.}} 
\label{Table 1}
%\linespread{0.5}

%\renewcommand{\footnoterule}{}
\centering
\begin{tabular}{cccccccc}
\hline
%%%
\hline
No. & ObsID & Start Time & Stop Time & CC  & Hardness Ratio & Branch Position on HID Plane\\ \\
\hline
%1 & & 2 & 3 & 4 & 5&  \\ \\
\hline
1	&10061-02-01-00		& 02-11-1996 08:34 	&  16:21  	& 	  0.65-0.96	     	&   0.46-0.54	     		& 	Lower and Upper NB \\
2	&10063-02-01-00		& 03-11-1996 20:30 	&  02:11  	&         0.83-0.95		   	&   0.42-0.45        	&    	Lower NB\\
3	&10061-02-02-000	& 06-11-1996 21:19 	&  05:13  	&         0.43-0.94		     	&    0.38-0.54	     	&    	Soft and Upper NB\\
4	&10061-02-02-00  	& 07-11-1996 05:13      &  05:34        &         0.85		       &    0.46-0.47	        &  	Lower and Middle NB\\
5	&10061-02-03-00		& 16-11-1996 00:55 	&  08:31  	&         0.85			     	&   0.60-0.67	     	&    	Lower and Upper HB\\
6	&20055-01-01-00		& 15-02-1997 08:32 	&  10:27  	&         -0.73		     	&    0.59-0.63	     		&    	Hard apex and Upper NB\\
7	&20055-01-02-00		& 12-04-1997 19:00 	&  20:55  	&         0.25-0.4		     	&    0.62-0.66	     		&    	Lower and Middle HB\\
8	&20055-01-03-00		& 29-05-1997 19:18 	&  21:55  	&         0.83 - -0.43	     	&    0.38-0.44	     		&    	Soft and Lower NB\\
9	&  20053-02-01-04	& 06-06-1997 00:28 	&  02:28  	&         0.42-0.73		&    0.48-0.56			&    	Upper NB\\
10	&  20053-02-02-00	& 25-07-1997 05:18 	&  10:00  	&         0.92			&    0.40-0.48	     		&    	Lower and Middle NB\\
11	&  20053-02-01-02	& 25-07-1997 11:43 	&  17:29  	&         0.25-0.65	     	&    0.45-0.65	     		&       Lower NB and Middle HB\\
12	&  20053-02-01-03	& 25-07-1997 18:23 	&  20:40  	&          0.35-0.8	     	&    0.63-0.67       		& 	Lower and Middle HB\\
13	&  20055-01-05-00	& 21-09-1997 12:29 	&  14:24  	&          0.32	   		&    0.56-0.59       		&    	Upper NB and Hard apex\\
14	&  30042-01-02-00	& 14-09-1998 00:25 	&  03:02  	&         0.30  		&    0.40-0.55	     		&    	Lower and Middle NB\\
15	&  30042-01-03-00	& 25-09-1998 03:28 	&  08:23  	&         -0.52			&    0.60-0.66	     		&    	Lower and Middle HB\\
16	&  30042-01-02-01	& 09-10-1998 03:20 	&  06:38  	&              0.98	    	&    0.40-0.53       		& 	Soft apex and Upper NB\\
17	&  30042-01-05-00	& 14-10-1998 00:16	&   05:59 	&         0.44-0.61		&    0.38-0.42	     		&    	Soft apex and Lower NB\\
18	&  30042-01-07-00	& 31-10-1998 05:23 	&  08:45  	&         0.92			&    0.40-0.45	     		&    	Soft apex and Lower NB\\
19	&  30042-01-08-01	& 02-11-1998 06:34 	&  08-27  	&         -0.85			&    0.74-0.78	     		&    	Upper HB\\
20	&  30042-01-08-00	& 03-11-1998 03:18 	&  09:04  	&      	0.47-0.64		&    0.62-0.66       		&    	Middle HB\\
21	&  30042-01-10-00	& 08-11-1998 06:36 	&  07:25  	&         -0.30		     	&    0.69-0.72			&    	Upper HB\\
22	&  30042-01-12-00	& 10-11-1998 00:07 	&  04:28  	&         0.14-0.35		     	&    0.66-0.72		&       Upper HB\\
23	&  30042-01-14-00	& 11-11-1998 03:16 	&  07:56  	&         0.27-0.44		     	&    0.65-0.71		&    	Upper HB\\
24	&  30042-01-17-00	& 21-11-1998 11:47 	&  18:28  	&         0.67-0.92		     	&    0.52-0.58	     	&    	Middle and Upper NB\\
25	&  30042-01-19-00	& 22-11-1998 06:35 	&  09:06  	&         0.42-0.58			&    0.52-0.59  	&    	Upper NB\\
26	&  30042-01-20-00	& 22-11-1998 09:47 	&  11:17  	&         0.73				&    0.51-0.54		&    	Upper NB\\
27	&  40018-02-01-000	& 01-03-2000 15:30 	&  23:30  	&         0.33-0.72		     	&    0.35-0.55	     	&    	FB and Upper NB\\
28      &  40018-02-01-00	& 01-03-2000 23:30 	&  03:43  	&         0.42-0.65		     	&    0.36-0.40	     	&    	FB and Soft apex\\
29	&  40018-02-01-10	& 02-03-2000 04:30 	&  05:08  	&         0.94			     	&    0.38-0.45	     	&    	FB and Lower NB\\
30	&  40018-02-01-03	& 02-03-2000 12:15 	&  12:59  	&         0.42		     		&    0.38-0.44       	&	FB and Lower NB\\
31	&  40018-02-01-01	& 02-03-2000 15:29 	&  16:41  	&          0.68    			&    0.36-0.41       	& 	FB and Lower NB\\
32	&  40018-02-01-04	& 02-03-2000 17:11 	&  00:01  	&         0.33-0.92			&    0.38-0.51	     	&    	FB and Middle NB\\
33	&  40018-02-02-10	& 03-03-2000 00:01 	&  03:37  	&         -0.63 		   	&    0.33-0.38	     	&    	FB and Soft apex\\
34	&  40018-02-02-02	& 03-03-2000 07:25 	&  13:04  	&         0.76-0.9		     	&    0.37-0.54	     	&    	FB and Upper NB\\
35	&  40018-02-02-03	& 03-03-2000 23:24 	&  03:39  	&         0.87-0.18  		&    0.38-0.46	     		&    	FB and Lower NB\\
36	&  40018-02-02-04	& 04-03-2000 05:57 	&  13:00  	&         0.42-0.95	     	&    0.38-0.52	     		&    	FB and Middle NB\\
37	&  40018-02-02-05	& 04-03-2000 13:52 	&  14:44  	&         0.84		     	&    0.36-0.44	     		&    	FB and Lower NB\\
38	&  40018-02-02-21	& 05-03-2000 00:58 	&  03:36  	&        0.22-0.43    	&    0.38-0.42       		& 	FB and Lower NB\\
39	&  40018-02-02-14	& 05-03-2000 04:15 	&  05:01  	&         0.42		     	&    0.34-0.37	     		&    	FB and Soft apex\\
40	&  40018-02-02-13	& 05-03-2000 17:56 	&  23:56  	&         0.24-0.91	     	&    0.35-0.55	     		&    	FB and Upper NB\\
41	&  40018-02-02-17	& 06-03-2000 02:34 	&  03:33  	&         0.25 	     		&    0.38-0.43	     	&    	FB and Lower NB\\
42	&  40018-02-02-22	& 06-03-2000 04:10 	&  05:09  	&         -0.43			    	&    0.40-0.42	     	&    	Lower NB\\
43	&  40018-02-02-15	& 06-03-2000 05:46 	&  06:44  	&         0.91			     	&    0.40-0.46	     	&    	Lower NB\\
44	&  40018-02-02-23	& 06-03-2000 09:08 	&  09:56  	&         0.85			     	&    0.42-0.55	     	&    	Lower and Upper NB\\
45	&  40018-02-02-16	& 06-03-2000 13:41 	&  14:40  	&         0.81			     	&    0.38-0.42	     	&    	FB and Lower NB\\
46	&  40018-02-01-08	& 06-03-2000 15:17 	&  16:29  	&         0.92			     	&    0.49-0.53	     	&    	Lower and Upper NB\\
47	&  40018-02-02-09	& 06-03-2000 16:53 	&  23:52  	&         0.55-0.85 		     	&    0.40-0.53	     	&    	Lower and Upper NB\\	
48	&  40018-02-02-19	& 07-03-2000 05:39 	&  09:53  	&         0.41-0.92		     	&    0.38-0.50	     	&    	FB and Middle NB\\
49	&  50017-02-01-00	& 18-07-2000 07:29 	&  10:21  	&         0.92			     	&    0.47-0.53	     	&    	Lower NB and Upper NB\\
50	&  50017-02-01-01	& 19-07-2000 10:57 	&  12:11  	&         0.54			     	&    0.38-0.40	     	&    	FB and Lower NB\\
51	&  80105-09-01-00	& 24-09-2003 15:53 	&  16:54  	&         0.25	  			&    0.41-0.43	     	&    	Lower NB\\
52	&  80105-09-01-01	& 24-09-2003 17:28 	&  18:28  	&         0.25		     		&    0.38-0.40	     	&    	FB and Soft apex\\
53	&  80105-09-01-02	& 24-09-2003 19:17 	&  19:53  	&         0.30	     			&    0.36-0.40       	& 	FB and Soft apex\\
54	&  90022-09-03-00	& 15-04-2004 12:42 	&  13:50  	&         0.32		     		&    0.77-0.79	     	&    	Upper HB\\
55	&  90022-09-07-01	& 17-04-2004 11:46 	&  13:00  	&         0.20		     		&    0.59-0.61	     	&    	Hard apex and Lower HB\\
56	&  90022-09-08-00	& 01-10-2004 04:34 	&  06:05  	&         0.60			     	&    0.41-0.44	     	&    	Lower NB\\
57	&  90022-09-08-04	& 02-10-2004 04:19 	&  04:55  	&         0.90			     	&    0.50-0.56	     	&    	Middle and Upper NB\\

\hline
\hline
\end{tabular}
\end{minipage}
\end{table}

\clearpage

\begin{table}
\begin{minipage}[t]
{\columnwidth}
\caption[solutions]{Log of observations for which anti-correlated hard and soft X-ray lags were detected 
(first four columns). 
The cross-correlation coefficients (CCs) along with their delays are shown in the fifth and sixth columns. 
The last two columns show hardness ratios and the corresponding branch positions on the Z-track, respectively.}
\label{Table 2}
%\linespread{1.0}
\tiny
\renewcommand{\footnoterule}{}
\centering
\begin{tabular}{cccccccccc}
\hline
%%%
\hline
No. & ObsID & Start Time & Stop Time & CC & Delay (s) & Hardness Ratio&Branch position on HID plane\\ \\
\hline
%1 & & 2 & 3 & 4 & 5&  \\ \\
\hline
1 & 20053-02-01-00&30-05-1997,	09:28	& 17:14	& a) -0.40$\pm$0.09  & a) 34.05$\pm$18.02 & 0.54 -- 0.56&Upper NB\\
& & & & 					  b) -0.59$\pm$0.06& b) 260.32$\pm$12.12& 0.53 -- 0.56&Upper NB\\

2 &20053-02-01-01&25-07-1997, 21:41&01:07&-0.60$\pm$0.06&	-249$\pm$108& 0.67 -- 0.74&Lower and Middle HB\\

3&20055-01-04-00&28-07-1997, 18:27&20:24&-0.72$\pm$0.05&	31.36$\pm$12.04& 0.52 -- 0.59&Upper NB\\

4&30042-01-01-00&22-08-1998, 10:36&16:31&-0.53$\pm$0.05&95.70$\pm$15.25&0.70 -- 0.73&Upper HB\\

5&30042-01-04-00&08-10-1998, 8:18&10:14	&-0.55$\pm$0.08&209.18$\pm$20.22&0.47 -- 0.58&Lower and Upper NB\\
6&30042-01-06-00&26-10-1998, 04:58&09:12& a) -0.41$\pm$0.04 & a) 580$\pm$142 &0.39 -- 0.44 & Soft apex\\
& & & &			b) -0.53$\pm$0.03& b) 748$\pm$62&0.69 -- 0.73&Upper HB\\

7&30042-01-09-00&04-11-1998, 05:01&08:15&-0.51$\pm$0.05	&150$\pm$21&0.69 -- 0.73&Upper HB\\
8&30042-01-11-00&09-11-1998, 06:38&10:45&a)-0.49$\pm$0.08&a)43$\pm$18&0.61 -- 0.65 & Lower HB\\
&&&&                    b)-0.88$\pm$0.05 & b)57$\pm$14 & 0.58 -- 0.64&Hard apex \& Lower HB\\
9&30042-01-13-00&10-11-1998, 06:39&12:21&a) -0.48$\pm$0.07  &a) 202$\pm$92& 0.70 -- 0.74&Upper HB\\
	& & & &				b)-0.61$\pm$0.09&b) 72$\pm$25&0.57 -- 0.60&Hard apex\\
10&30042-01-15-00	&20-11-1998, 16:10&18:34	&a) -0.59$\pm$0.05&a) 748$\pm$51&0.58 -- 0.66 & Hard apex and Lower HB\\
	& & & &				b) -0.78$\pm$0.04&b) 58$\pm$8&0.57 -- 0.62 & Hard apex\\
11&30042-01-16-000	&21-11-1998, 00:10&08:03	&-0.57$\pm$0.04&67$\pm$14&0.38 -- 0.40 & FB\\
12&30042-01-16-00&	21-11-1998, 08:37&10:47	&-0.59$\pm$0.02	&220$\pm$32& 0.37 -- 0.42&FB and Soft apex\\
13&30042-01-18-00	&21-11-1998, 22:35&05:14& a) -0.42$\pm$0.06 &a) -160$\pm$70 &0.58 -- 0.61 & Lower HB, Hard apex and Upper NB\\
		& & & &				b) -0.61$\pm$0.08&b) 65$\pm$20&0.60 -- 0.61 & Lower HB\\
14&40018-02-02-030 	&03-03-2000, 15:25&23:24&-0.51$\pm$0.03&284$\pm$13&0.40--0.52&Lower NB \\
15&40018-02-02-210	&04-03-2000, 17:04&00:58&-0.58$\pm$0.02&-920$\pm$102&0.38--0.50 &FB and Lower NB\\
16&90022-09-01-00	&02-04-2003, 02:59&02:36&-0.60$\pm$0.05&-256$\pm$102&0.57--0.60&Hard apex\\
17&90022-09-07-00&17-04-2004, 06:45&07:59& -0.39$\pm$0.08&	180$\pm$30&0.58--0.60&Hard apex\\
\hline
\hline
\end{tabular}
\end{minipage}
\end{table}

\clearpage

\begin{table}
\begin{minipage}[t]{\columnwidth}
\scriptsize
\caption{Best-fit spectral parameters for the ObsID 30042-01-13-00. The letters A, B, C, and D represent 
the segments in the light curve. The subscripts dbb and bb represent the disk black body and black body model, 
respectively. 
The flux in units of 10$^{-8}$ ergs cm$^{-2}$ s$^{-1}$ is calculated in the energy band 3--25 keV
and otherwise it is mentioned.
Errors are quoted at a 90\% confidence level.} 
\label{tab1}
\centering
\renewcommand{\footnoterule}{}
\begin{tabular}{ccccccccc}
\hline
\hline
Parameters&\multicolumn{2}{c}{A}&\multicolumn{2}{c}{B}&\multicolumn{2}{c}{C}&\multicolumn{2}{c}{D}\\
\hline
&A$_{dbb}$&A$_{bb}$&B$_{dbb}$&B$_{bb}$&C$_{dbb}$&C$_{bb}$&D$_{dbb}$&D$_{bb}$\\
 %&diskbb&bb&diskbb&bb\\
\hline
\hline

$kT_{in}$ (keV)\footnote{Inner disk temperature using the disk black body model.}&2.10$\pm$0.07 &-&2.36$\pm$0.08&-&2.82$\pm$0.03&-&2.87$\pm$0.03&-\\
$N_{disk}$\footnote{Normalization of the disk black body model.}&52.52$\pm$5.50 &-&25.94$\pm$2.26&-&16.54$\pm$1.80&-&16.25$\pm$2.01&-\\

$kT_{bb}$ (keV)\footnote{Temperature of the black body model.}&-&1.51$\pm$0.10 &- &1.67$\pm$0.11&-&1.96$\pm$0.05&-&2.04$\pm$0.05\\
$N_{bb}$\footnote{Normalization of the black body model.}& -&0.09$\pm$0.01 &-&0.06$\pm$0.01&-&0.07$\pm$0.02&-&0.08$\pm$0.02\\
kT$_{e}$\footnote{Electron temperature.}(keV)& 3.3$\pm$0.2&3.00$\pm$0.01&3.3$\pm$0.3&3.0$\pm$0.1&6.1$\pm$1.3&3.5$\pm$0.3&12.8$\pm$2.3&3.7$\pm$0.4\\
$\tau$\footnote{Optical depth of the Compton cloud.}& 8.7$\pm$0.6&9.3$\pm$0.5&8.6$\pm$0.8&9.8$\pm$0.6&4.3$\pm$0.7&8.4$\pm$0.8&2.2$\pm$1.2&8.1$\pm$0.9\\
disk flux& 1.54&-&1.22&-&1.60&-&1.72&-\\
CompTT flux&2.27&-&2.20&-&1.19&-&1.00&-\\
BB flux\footnote{in units of 10$^{-9}$ ergs cm$^{-2}$ s$^{-1}$.}&-&6.14 &-&4.34 &-&5.89&-&6.19\\
CompTT flux&-&3.06 &-&3.07 &-&2.26&-&2.14\\
Delay (s)&202$\pm$92&-&-&-&139$\pm$36&-&72$\pm$25&-\\
pivot energy (keV)&6.7$\pm$0.1&-&-&-&12.2$\pm$0.2&-&15.6$\pm$0.3&-\\
$\nu_{centriod}$\footnote{HBO.}(Hz)&51.7$\pm$1.8&-&39.1$\pm$1.0&-& 24.2$\pm$0.2&-&21.5$\pm$0.3&-\\
$\chi^{2}$/dof&23/46 & 25/46&25/46&23/46&28/46&27/46&26/46&27/46\\

\hline
\hline
\end{tabular}
\end{minipage}
\end{table}

%\clearpage
\begin{table}
\begin{minipage}[t]{\columnwidth}
\caption{Best-fit spectral parameters for the spectra of A and B segments for ObsID 20053-02-01-01. 
Errors are quoted at a 90\% confidence level.} 
\label{tab1}
\centering
\renewcommand{\footnoterule}{}
\begin{tabular}{ccccccccc}
\hline
\hline
Parameters&\multicolumn{2}{c}{20053-02-01-01}\\
\hline
&A &B\\
 %&diskbb&bb&diskbb&bb\\
\hline
\hline

$kT_{in}$ (keV)\footnote{Inner disk temperature using the disk black body model.}&2.84$\pm$0.01 &2.95$\pm$0.04 \\
$N_{disk}$\footnote{Normalization of the disk black body model.}&16.7$\pm$0.3 &14.6$\pm$0.1\\

kT$_{e}$\footnote{Electron temperature.}(keV)& 7.7$\pm$0.2&13.1$\pm$0.3\\
$\tau$\footnote{Optical depth of the Compton cloud.}& 3.5$\pm$0.7 & 2.0$\pm$0.5  \\
Delay (s)&-249$\pm$108&-\\
pivot energy (keV)&8.9$\pm$0.1&-\\
$\nu_{centriod}$\footnote{HBO.}(Hz)&23.1$\pm$0.1&18.2$\pm$0.1\\
kHz $\nu_{centriod}$(Hz)&588$\pm$28&546$\pm$8\\
$\chi^{2}$/dof&31/47 & 36/47&\\

\hline
\hline
\end{tabular}
\end{minipage}
\end{table}

\label{lastpage}

\end{document}